\documentclass[12pt]{article}

\usepackage{epsfig}

\def\hybrid{\topmargin -20pt  \oddsidemargin 0pt
      \headheight 0pt   \headsep 0pt
      \textwidth 6.25in 
      \textheight 9.5in 
      \marginparwidth .875in
      \parskip 5pt plus 1pt   \jot = 1.5ex}

\hybrid

\let\LARGE=\large

\let\large=\normalsize

\def\pid{(2 \pi)^3}

\begin{document}
\def\x{\times}
\def\beq{\begin{equation}}
\def\eeq{\end{equation}}
\def\beqa{\begin{eqnarray}}
\def\eeqa{\end{eqnarray}}
\def\L{ {\cal L}}
\def\C{ {\cal C}}
\def\N{ {\cal N}}
\def\dd{{\rm d}}
\def\Re{{\rm Re}}
\def\Im{{\rm Im}}
\def\calE{{\cal E}}
\def\lin{{\rm lin}}
\def\Tr{{\rm Tr}}
\def\cF{{\cal F}}
\def\cD{{\cal D}}
\def\modS{{S+\bar S}}
\def\mods{{s+\bar s}}
\newcommand{\Fg}[1]{{F}^{({#1})}}
\newcommand{\cFg}[1]{{\cal F}^{({#1})}}
\newcommand{\cFgc}[1]{{\cal F}^{({#1})\,{\rm cov}}}
\newcommand{\Fgc}[1]{{F}^{({#1})\,{\rm cov}}}
\def\mpl{m_{\rm Planck}}
\def\mxth{\mathsurround=0pt }
\def\xversim#1#2{\lower2.pt\vbox{\baselineskip0pt \lineskip-.5pt
x  \ialign{$\mxth#1\hfil##\hfil$\crcr#2\crcr\sim\crcr}}}
\def\simgr{\mathrel{\mathpalette\xversim >}}
\def\simle{\mathrel{\mathpalette\xversim <}}

\newcommand{\ms}[1]{\mbox{\scriptsize #1}}
\renewcommand{\a}{\alpha}
\renewcommand{\b}{\beta}
\renewcommand{\c}{\gamma}
\renewcommand{\d}{\delta}
\newcommand{\pa}{\partial}
\newcommand{\g}{\gamma}
\newcommand{\G}{\Gamma}
\newcommand{\A}{\Alpha}
\newcommand{\B}{\Beta}
\newcommand{\D}{\Delta}
\newcommand{\e}{\epsilon}
\newcommand{\E}{\Epsilon}
\newcommand{\z}{\zeta}
\newcommand{\Z}{\Zeta}
\newcommand{\K}{\Kappa}
\renewcommand{\l}{\lambda}
\renewcommand{\L}{\Lambda}
\newcommand{\m}{\mu}
\newcommand{\M}{\Mu}
\newcommand{\n}{\nu}
\newcommand{\X}{\Chi}
\newcommand{\R}{\Rho}
\newcommand{\s}{\sigma}
\renewcommand{\S}{\Sigma}
\renewcommand{\t}{\tau}
\newcommand{\T}{\Tau}
\newcommand{\y}{\upsilon}
\newcommand{\Y}{\upsilon}
\renewcommand{\o}{\omega}
\newcommand{\q}{\theta}
\newcommand{\h}{\eta}

\def\cs#1{\footnote{{\bf Stefan:~}#1}}
\def\cd#1{\footnote{{\bf Dieter:~}#1}}
\def\cj#1{\footnote{{\bf Jan:~}#1}}
\def\cg#1{\footnote{{\bf Gottfried:~}#1}}
\def\ca#1{\footnote{{\bf Albrecht:~}#1}}

\def\dota{ {\dot{\alpha}} }
\def\lag{Lagrangian}
\def\Kahler{K\"{a}hler}
\def\kahler{K\"{a}hler}
\def\A{ {\cal A}}
\def\C{ {\cal C}}
\def\F{{\cal F}}
\def\cL{ {\cal L}}

\def\R{ {\cal R}}
\def\x{ \times }
\def\beq{\begin{equation}}
\def\eeq{\end{equation}}
\def\beqa{\begin{eqnarray}}
\def\eeqa{\end{eqnarray}}

\sloppy
\newcommand{\be}{\begin{equation}}
\newcommand{\eq}{\end{equation}}
\newcommand{\ov}{\overline}
\newcommand{\un}{\underline}
\newcommand{\p}{\partial}
\newcommand{\la}{\langle}
\newcommand{\ra}{\rangle}
\newcommand{\bl}{\boldmath}
\newcommand{\ds}{\displaystyle}
\newcommand{\nl}{\newline}
\newcommand{\Nzahl}{{\bf N}  }
\newcommand{\zzahl}{ {\bf Z} }
\newcommand{\Zzahl}{ {\bf Z} }
\newcommand{\Qzahl}{ {\bf Q}  }
\newcommand{\Rzahl}{ {\bf R} }
\newcommand{\Czahl}{ {\bf C} }
\newcommand{\wt}{\widetilde}
\newcommand{\wh}{\widehat}
\newcommand{\fs}[1]{\mbox{\scriptsize \bf #1}}
\newcommand{\ft}[1]{\mbox{\tiny \bf #1}}
\newtheorem{satz}{Satz}[section]
\newenvironment{Satz}{\begin{satz} \sf}{\end{satz}}
\newtheorem{definition}{Definition}[section]
\newenvironment{Definition}{\begin{definition} \rm}{\end{definition}}
\newtheorem{bem}{Bemerkung}
\newenvironment{Bem}{\begin{bem} \rm}{\end{bem}}
\newtheorem{bsp}{Beispiel}
\newenvironment{Bsp}{\begin{bsp} \rm}{\end{bsp}}
\renewcommand{\arraystretch}{1.5}



\renewcommand{\thesection}{\arabic{section}}
\renewcommand{\theequation}{\thesection.\arabic{equation}}

\parindent0em

\def\S4{\frac{SO(4,2)}{SO(4) \otimes SO(2)}}
\def\P3{\frac{SO(3,2)}{SO(3) \otimes SO(2)}}
\def\MGd{\frac{SO(r,p)}{SO(r) \otimes SO(p)}}
\def\SOd{\frac{SO(r,2)}{SO(r) \otimes SO(2)}}
\def\SO2{\frac{SO(2,2)}{SO(2) \otimes SO(2)}}
\def\SUm{\frac{SU(n,m)}{SU(n) \otimes SU(m) \otimes U(1)}}
\def\SUS{\frac{SU(n,1)}{SU(n) \otimes U(1)}}
\def\SK{\frac{SU(2,1)}{SU(2) \otimes U(1)}}
\def\SU{\frac{ SU(1,1)}{U(1)}}

\begin{titlepage}
\begin{center}
\hfill HU-EP-00/58\\
\hfill AEI-2000-84\\
\hfill {\tt hep-th/0012213}\\

\vskip .1in

{\LARGE {\bf On the Vacuum Structure of Type II String Compactifications
on Calabi-Yau Spaces with H-Fluxes}}

\vskip .2in

{\bf Gottfried Curio$^a$, Albrecht Klemm$^a$,  
Dieter L\"ust$^a$ and Stefan Theisen$^b$}\footnote{\mbox{
email: \tt 
aklemm,curio,luest@physik.hu-berlin.de, theisen@aei-potsdam.mpg.de}}
\\
\vskip 1cm

$^a${\em Humboldt-Universit\"at zu Berlin,
Institut f\"ur Physik, 
D-10115 Berlin, Germany}\\
$^b${\em Max-Planck-Institut f\"ur Gravitationsphysik, D-14476 Golm,
}

\vskip .1in

\end{center}

\vskip .2in

\begin{center} {\bf ABSTRACT} \end{center}
\noindent
We discuss the vacuum structure of type IIA/B
Calabi-Yau string compactifications to four dimensions in the presence
of $n$-form H-fluxes. These will lift the vacuum degeneracy in the
Calabi-Yau moduli space, and for generic points in the moduli space,
${\cal N}=2$ supersymmetry will be broken. 
However, for certain `aligned' choices of the H-flux vector, supersymmetric 
ground states are possible at the degeneration points of the Calabi-Yau 
geometry. We will investigate in detail the H-flux induced superpotential  
and the corresponding scalar potential at several degeneration points, 
such as the Calabi-Yau large volume limit, the conifold loci, the Seiberg-Witten points, the 
strong coupling point and the conformal points.
Some emphasis is given to the question whether partial supersymmetry
breaking can be realized at those points. 
We also relate the H-flux induced superpotential to the
formalism of gauged ${\cal N}=2$ supergravity.
Finally we point out the analogies between the Calabi-Yau vacuum structure
due to H-fluxes and the attractor formalism of 
${\cal N}=2$ black holes.
\vskip .5cm
\noindent
December 2000\\
\end{titlepage}
\vfill
\eject

\newpage

\section{Introduction}

In this paper we will discuss the vacuum structure of type II strings
on Calabi-Yau three-folds with internal $n$-form H-fluxes  turned on.
In general, the effect of non-vanishing H-fluxes is that they
lift the vacuum degeneracy in the Calabi-Yau moduli space. In fact,
as already discussed in \cite{PolStro,Mich,TayVa,Mayr}, at generic points
in the Calabi-Yau moduli space,
non-trivial Ramond and/or
Neveu-Schwarz $n$-form H-fluxes generally 
break ${\cal N}=2$ space-time supersymmetry completely, unless
their contribution to the vacuum energy is  balanced by other background
fields, such as the dilaton field in heterotic string compactifications 
\cite{warp}.
However ${\cal N}=2$ or ${\cal N}=1$ supersymmetric vacua  can be found
at certain corners in the  moduli space, where the Calabi-Yau
geometry is degenerate. We will consider
several degeneration points of the Calabi-Yau geometry, such as
the large volume limit, the Calabi-Yau conifold point, the Seiberg-Witten limit
and the strong coupling singularity. However as
soon as one abandons these special points, supersymmetry will be in general
broken. E.g. going away from the classical large radius limit, 
type IIA world-sheet instanton corrections to the prepotential
imply a non-degenerate period vector such that supersymmetry gets
broken \cite{Mayr}.
In addition, there might be the possibility
for unbroken supersymmetry in case
the contribution of the Ramond fluxes is balanced by the
NS-fluxes, as we will discuss at the end of the paper.

Turning on $n$-form H-fluxes 
on the six-dimensional Calabi-Yau space  
corresponds to a gauging of certain hypermultiplet isometries in the
low-energy ${\cal N}=2$ supergravity action and 
leads 
to a non-vanishing scalar potential 
in four dimensions which lifts the previous vacuum 
degeneracy \cite{PolStro,Mich}. Alternatively the H-fluxes can be described
by
a non-trivial superpotential
$W$ in four dimensions,
which is expressed in terms of ${\cal N}=1$ chiral fields \cite{TayVa}. 
Specifically, it turns
out that the superpotential is simply given by the 
symplectic scalar product
of the (dilaton dependent)  H-flux vector with the ${\cal N}=2$
period vector $\Xi$, 
which is a function of the complex scalars residing in the
${\cal N}=2$ vector multiplets.\footnote{Related types of Calabi-Yau
superpotentials were discussed before in \cite{FKLZ}.} In this way
the superpotential is closely tied up to the Calabi-Yau geometry,
since the period vector $\Xi$ corresponds to the various geometric
cycles of the Calabi-Yau space.
The question of supersymmetry breaking is then intimately related to
the question whether the H-fluxes are turned on in the directions
of the vanishing cycles of the Calabi-Yau spaces (aligned case) or not
(misaligned case). For the aligned situations, the degeneration
points in the Calabi-Yau geometry are attractor points where supersymmetry
is unbroken and the potential exhibits a (local) minimum of zero energy. 
On the other hand, in case of H-fluxes which are misaligned with
respect to a particular vanishing cycle, supersymmetry will be broken 
at the degeneration points in the Calabi-Yau geometry. Therefore
the question which degeneration point corresponds to a supersymmetric
ground state depends crucially on the chosen H-fluxes.

In this paper we will first show that the gauging of ${\cal N}=2$
supergravity due to H-fluxes leads to the superpotential of 
\cite{TayVa}. Subsequently we will
discuss in detail the vacuum structure of type II Calabi-Yau
compactifications with H-flux induced superpotential.
The paper is organized as follows. In the next section we shortly review
those aspects of ${\cal N}=2$ special geometry, 
which we need for our discussion, as well as the derivation 
of the symplectic invariant 
superpotential from the gauged hypermultiplet couplings.
Analyzing the structure of the gravitino mass matrix which follows
from the H-flux induced superpotential we will 
see that partial supersymmetry breaking from ${\cal N}=2$ to
${\cal N}=1$ supersymmetry \cite{APT,FGP}
is a priori possible in case the flux
vector is complex which means that Ramond as well as NS fluxes have
to be turned on. However treating the type IIB dilaton field as a
dynamical variable 
${\cal N}=2$ supersymmetry will be either unbroken
or completely broken at the minimum of the scalar potential. Therefore 
at the degeneration
points with aligned fluxes in the Calabi-Yau geometry, 
the full ${\cal N}=2$ supersymmetry is unbroken. 
Finally, at the end of sect. 2,
we point out that the superpotential formalism due to internal H-fluxes is
closely related to ${\cal N}=2$ black holes and the so called attractor
formalism \cite{FerKal1,moore}. 
In fact, when computing the supersymmetric points in the effective
supergravity action one has to solve precisely the same 
equations which determine the scalar fields at the horizon of the
${\cal N}=2$ black holes.
This means that the supersymmetric ground states with H-fluxes are
the attractor points of the ${\cal N}=2$ black holes.

In sect. 3 we discuss 
in detail the vacuum structure of type II compactifications with
non-trivial Ramond H-fluxes turned on. We focus on the special points
in the Calabi-Yau moduli spaces where the H-fluxes are aligned
with the vanishing cycles. We will see that the correct
identification of the vanishing cycles might be quite subtle, as in the case
of the Seiberg-Witten limit.
We should note that while our discussion will concentrate on specific
Calabi-Yau compactifications, qualitatively the results will be generic, 
i.e. they are also valid for compactifications on other CY manifolds 
with the same type of special points in their moduli space. 

In chapter 4 we discuss changes of the above scenarios
in case NS H-fluxes are turned on. Studying one simple example we see
that supersymmetric vacua might be possible
away from the special points discussed before.

In the appendix we give additional details about the minimization of the 
potential in the perturbative heterotic limit with general flux vectors.

Related issues of H-fluxes in M-theory and type II compactifications on Calabi-Yau
fourfolds were discussed in \cite{gukov} and  \cite{HLM}, respectively.

\section{The superpotential from H-fluxes}

\setcounter{equation}{0}

\subsection{Special geometry and vector couplings}

The self-couplings of (Abelian) vector multiplets of ${\cal N}=2$ 
supersymmetric Yang-Mills theory 
are completely specified by a holomorphic function $F(X)$ 
of the complex scalar components of the $N_V$    
vector multiplets. With local supersymmetry this function 
depends on one additional, unphysical 
scalar field, which incorporates the 
graviphoton. 
Including this field, the Abelian gauge group is $G=U(1)^{N_V+1}$.
The couplings of the vectors 
are now encoded in a function $F(X)$ of the complex scalars
$X^I$, $I=0,\dots,N_V$. $F(X)$ is holomorphic and homogeneous
of degree two. 

More abstractly, 
the {\em special geometry} \cite{DWVP,special} of the 
K\"ahler manifold ${\cal M}$ parameterized by the $N_V$ physical scalars 
is defined in terms of $2(N_V+1)$ 
covariantly holomorphic sections $L^I,\,M_I$ of a bundle ${\cal L}\otimes {\cal V}$ where 
${\cal L}$ is a line bundle and ${\cal V}$ is a $Sp(2(N_V+1),{\bf Z})$-bundle;   
i.e. $D_{\bar A} L^I
=(\partial_{\bar A}-{1\over2}\partial_{\bar A} K_V(z,\bar z))L^I=0$, 
and likewise for the $M_I$. Here $K_V(z,\bar z)$ is the 
K\"ahler potential, and the physical scalars $z^A,\,A=1,\dots, N_V$ are 
intrinsic complex coordinates on the moduli space ${\cal M}_V={\cal SK}(N_V)$,
which is a special K\"ahler space of complex dimension $N_V$.   
The sections are assembled into a symplectic vector $V$:
\beqa
V= \pmatrix{ L^I \cr M_I\cr} \;\;, \label{sympsection}
\eeqa
${\cal M}_V$ is defined by the constraint 
\beqa
\langle \bar V,V\rangle \equiv \bar V^{\rm T}\Omega V 
= -i ,\label{symconstr}
\eeqa 
with $\Omega$ the invariant symplectic metric
\beqa
\Omega = \pmatrix{ 0& {\bf 1} \cr -{\bf 1} &0 \cr }\,.
\eeqa
Given $X^I$ and $F_I$, one may define an  holomorphic period vector $\Xi$,
\beqa
\Xi(z)= \pmatrix{ X^I(z) \cr F_I(z)\cr} \;\;, \label{sympsectionhol}
\eeqa
 via 
\beq
X^I(z) = e^{-{1\over 2}K_V(z,\bar z)}\,L^I\,,\quad
F_I(z) = e^{-{1\over 2}K_V(z,\bar z)}\,M_I\,,
\label{section} 
\eeq

Via the constraint (\ref{symconstr}) the K\"ahler potential 
can be expressed as
\begin{equation}
K_V(z,\bar z)=
-\log\Big(i \bar X^I(\bar z)F_I(z)-i X^I(z)
\bar F_I(\bar z)\Big)= - \log(i \Xi^{\dagger} \Omega \Xi)  \  .
\label{KP} 
\end{equation}
Invariance under $Sp(2 N_V+2)$ transformations is manifest. 

If $\det(\partial_i X^I,X^I)\neq 0$,
there exists a holomorphic, homogeneous prepotential $F(X)$ such that 
$F_I=\partial F(X)/\partial X^I$.
In this case the $X^I$ are good local homogeneous coordinates
on ${\cal M}_V$; they are algebraically independent, 
i.e. $\partial_I X^J=\delta_I^J$.
The existence of a prepotential is a basis-dependent statement. 
There exists, however, always a symplectic basis $(X^I,F_I)$ such that 
$F_I=\partial_I F(X)$.

One important example which after a symplectic 
transformation leads to algebraically dependent periods is given
by a prepotential which is linear in one of the sections, say
in $X^1$:
\begin{equation}
F(X)= t^1 G(X^0,X^a)+H(X^0,X^a).\label{depperiods}
\end{equation}
Here we have defined $t^1=X^1/X^0$ and $G$, $H$ are functions of  
$X^0$ and $X^a$ ($a=2,\dots ,N_V$) only.
Then $F_1=G/X^0$ is independent of $X^1$, and after the symplectic 
transformation $X^1\rightarrow\tilde X^1=F_1$ the new $\tilde X^I$ are
algebraically dependent.

If a prepotential exists for the basis $(X^I,F_I)$, 
we can introduce inhomogeneous coordinates $t^A$ on ${\cal M}_V$
which are defined as\footnote{Later we will also use inhomogeneous coordinates
$T^A=-i{X^A\over X^0}$.} 
\begin{equation}
t^A={X^A\over X^0},\quad X^0\neq 0\,,\quad A=1,\ldots, N_V.
\end{equation}
In this parameterization the K\"ahler potential is 
\cite{SU} 
\begin{equation}
K_V(t,\bar t) = -\log\Big(2({\cal F}+ \bar{\cal F})-
           (t^A-\bar t^A)({\cal F}_A-\bar{\cal F}_A)\Big)\,,
\label{Kspecial}
\end{equation}
where ${\cal F}(z)=i(X^0)^{-2}F(X)$.

\subsection{Hypermultiplet Couplings and superpotential}

\subsubsection{Gauged ${\cal N}=2$ supergravity}

Now consider the ${\cal N}=2$ supergravity couplings including
$N_H$ hypermultiplets $q_i$ as additional matter fields \cite{andrianop}.
Together with the $N_V$ vector multiplets the moduli space is locally, at generic 
points in the moduli space, a product space of the form
\begin{equation}
{\cal M}={\cal SK}(N_V)\otimes {\cal Q}(N_H),
\end{equation}
where the hypermultiplet moduli
space ${\cal M}_H={\cal Q}(N_H)$ 
is a quaternionic space of real dimension $4N_H$.
The coupling of the hypermultiplet scalars $q_i$ to the vectormultiplets $X^I$
arises from gauging the Abelian isometries of ${\cal Q}$. This
means that the hypermultiplets are charged with respect to the
gauge group $G=U(1)^{N_V+1}$. 
The gauging is done by introducing $N_V+1$ Killing vectors $k^i_I(q)$ 
on ${\cal M}_H$
which correspond to the (field dependent) 
Abelian charges of the hypermultiplets.
This means that one defines the following covariant derivatives
\begin{equation}
\nabla_\mu q^i=\partial_\mu q^i+k^i_I A^I_\mu\, .
\end{equation}
The Killing vectors $k^i_I$ can be determined in terms of 
a $SU(2)$  triplet of real Killing prepotentials $P^x_I$ as follows
\begin{equation}
k^i_I\Omega^x_{ij}=-(\partial_j P^x_I+\epsilon^{zyz}\omega^y_jP^z_I)\, ,
\end{equation}
where $\omega^x$ is a $SU(2)$ connection and $\Omega^x$ its 
curvature.

The gauging of the hypermultiplet isometries generically implies 
non-vanishing masses of the two ${\cal N}=2$ gravitini, whose mass
matrix is:
\begin{equation}
S_{AB}={i\over 2}e^{K_V/2}~(\sigma_x)_A{}^C~\epsilon_{BC}~P^x_I(q)~X^I(z)
={i\over 2}(\sigma_x)_A{}^C~\epsilon_{BC}~P^x_I(q)~L^I\, .
\label{gravmat}
\end{equation}

We now rewrite the coupling of the
${\cal N}=2$ hypermultiplets to the vector multiplets in 
${\cal N}=1$ language. 
The coupling with $x=3$ corresponds to ${\cal N}=1$
D-terms while those with $x=1,2$ to F-terms, 
i.e. they are equivalent to a ${\cal N}=1$ superpotential.

{}From now on we are interested in situations where all  $P^3_I=0$.
To derive the superpotential let us introduce the following 
functions $e_I$:
\begin{equation}
e_I=e^{-K_H/2}(P^1_I+iP^2_I)\, .\label{eii}
\end{equation}
Then the ${\cal N}=1$ superpotential is
\begin{equation}
W(z,q)=e_I(q)X^I(z),\label{supo}
\end{equation}
where the $z^A$ and also the $q_i$ now denote ${\cal N}=1$
chiral superfields.
We will now motivate (\ref{supo}).

The ${\cal N}=1$ supergravity action can
be expressed in terms of the generalized K\"ahler function
\begin{equation}
G=K_V(z,\bar z)+K_H(q,\bar q)+\log|W(z,q)|^2.
\end{equation}
The scalar potential is
\begin{equation}
v=e^G\biggl( G_{A}G_{\bar B}G^{A\bar B}+G_{i}G_{\bar\jmath}
G^{i\bar\jmath}-3\biggr) ,\label{scalarpot}
\end{equation}
and the ${\cal N}=1$ gravitino mass  
\begin{equation}
m_{3/2}=e^{G/2}=e^{K/2}|W|.\label{gravmass}
\end{equation}

The introduction of the superpotential eq.(\ref{supo}) is largely based on
the fact that the mass of the ${\cal N}=1$ gravitino 
in eq.(\ref{gravmass}) agrees with one of the two  mass
eigenvalues of the two ${\cal N}=2$ gravitini in eq.(\ref{gravmat}):
\begin{eqnarray}
S_{AB}={i\over 2}e^{(K_V+K_H)/2}\left(\matrix{
      -W+2i~\Im(e_I) ~X^I   &      0     \cr 
  0     &      W }
\right)\label{gravmatb}
\end{eqnarray}
Indeed, one eigenvalue agrees with (\ref{gravmass}). 
However for complex $e_I$ and $m^I$ the mass of the second
gravitino is in general different.

\subsubsection{Symplectic covariance}

Since the superpotential eq.(\ref{supo}) only contains the periods
$X^I$ but not the dual periods $F_I$, it is clear that the gravitino
masses so far are 
not invariant under symplectic $Sp(2N_V+2,{\bf Z})$ transformations. 
One can achieve full symplectic invariance by introducing 
magnetic prepotentials $\tilde P^{xI}$ \cite{Mich}.
These can be thought of as giving the relevant hypermultiplets also
a magnetic charge with respect to the Abelian gauge group $U(1)^{N_V+1}$,
which can be done by introducing magnetic Killing vectors $\tilde k^{iI}$. 
It then follows that  the electric/magnetic prepotentials
$(P_I^x,\tilde P^{xI})$ as well as the corresponding Killing vectors
$(k_I^i,\tilde k^{iI})$  transform as vectors under
$Sp(2N_V+2)$.

In analogy with the $e_I$ in eq.(\ref{eii}) 
we introduce  the complex magnetic functions 
\begin{equation}
m^I=e^{-K_H/2}(\tilde P^{1I}+i\tilde P^{2I})\, .
\end{equation}
Then the $e_I$ and 
the $m^I$ build a symplectic vector $H$ of the form
\begin{equation}
H=\pmatrix{m^I\cr e_I\cr}\, ,
\end{equation}
and the superpotential is given by the symplectic invariant 
scalar product between $H$ and $\Xi$:
\begin{equation}
W(z,q)=\langle H,\Xi\rangle=
e_I(q)X^I(z)-m^I(q)F_I(X^I(z)),\label{supom}
\end{equation}
The  ${\cal N}=1$
gravitino mass can be simply
written as:
\begin{equation}
m_{3/2}=e^{(K_V+K_H)/2}|\langle H,\Xi\rangle |\, .
\end{equation}
Finally, the symplectic invariant ${\cal N}=2$  gravitino mass matrix is
\begin{eqnarray}
S_{AB}={i\over 2}e^{(K_V+K_H)/2}\left(\matrix{
 W+2i~\Im(e_I) ~X^I+2i~ \Im(m^I)~F_I    &      0     \cr 
  0     &      W }
\right)\label{gravmatbb}
\end{eqnarray}

Actually, the symplectic transformations act on the two vectors 
$(P_I^1,\tilde P^{1I})=e^{K_H/2}(\Re ~e_I,\Re ~ m^I)$ 
and $(P_I^2,\tilde P^{2I})=e^{K_H/2}(\Im ~ e_I,\Im ~ m^I)$.
This means that one can always perform symplectic transformations
such that, e.g., $(P_I^1,\tilde P^{1I})$ is purely electric. In addition,  
$(P_I^2,\tilde P^{I2})$ can be also made purely electric by a further
symplectic transformations in case these two vectors are local w.r.t.  each
other, i.e. if
\begin{equation}
\tilde P ~\times ~P\equiv P_I^1\tilde P^{2I}-\tilde P^{1I} P_I^2=0\, .
\label{aaa}
\end{equation}
If (\ref{aaa}) holds the superpotential  
can be always brought to the form eq.(\ref{supo}).
On the other hand, if $\tilde P ~\times ~P\neq 0$,
i.e. if there exist hypermultiplets with mutually non-local electric/magnetic
$U(1)$ charges, the superpotential necessarily contains both
$X^I$ and $F_I$ fields in any basis.  
Of course, it is not possible to write
down a Lorentz invariant microscopic gauged ${\cal N}=2$
supergravity action which contains hypermultiplets with mutually non-local 
electric/magnetic charges. But this case will be of interest
for us below, when we investigate points in the Calabi-Yau
moduli spaces where electric and dual (magnetic) cycles,
which mutually intersect, degenerate (Argyres Douglas points). 
Here the superpotential will contain both $X^I$ and $F_I$. 
We will assume that while a local action without additional 
auxiliary degrees of freedom does not exist 
for the non-local Argyres-Douglas points,
the effective superpotential and the 
corresponding gravitino mass formulas do provide a valid 
description for the massless fields.

\subsubsection{The ground state of the theory -- the question of
partial supersymmetry breaking}

The ground state of the theory is determined by the requirement that
the scalar potential is minimized with respect to all scalar fields:
\begin{equation}
{dv\over dz^A}=0,\quad {dv\over dq^i}=0 
\quad\rightarrow\quad  z^A=z^A|_{\rm min},\quad
q^i=q^i|_{\rm min}.
\end{equation}
${\cal N}=1$ supersymmetry is unbroken at the minimum of the potential
if the auxiliary fields are zero
\begin{eqnarray}
h^{\bar A} &=& G^{\bar A B} e^{G/2} \partial_{B} G= G^{\bar A B} 
|W|~e^{K/2}~ \biggl(\partial_B K + {1\over W} \partial_B W \biggr)=0,\nonumber\\
h^{\bar \imath} &=& G^{\bar \imath j} e^{G/2} \partial_{j} G= G^{\bar \imath j} 
|W|~e^{K/2}~ \biggl(\partial_j K + {1\over W} \partial_j W \biggr)=0.
\label{auxfields}
\end{eqnarray}
In supergravity, supersymmetric minima of $v$ generally 
lead to a negative vacuum energy.
In order to find minima of $v$ with $v|_{\rm min}=0$ plus 
unbroken ${\cal N}=1$ supersymmetry, all four terms in eqs.(\ref{auxfields})
like $G^{A\bar B}|W|e^{K/2}\partial_BK$ etc.
must be separately zero. If $G^{\bar A B}e^{K/2}\partial_BK$,
$G^{\bar \imath j}e^{K/2}\partial_jK$,
and $G^{\bar A B}e^{K/2}$, $G^{\bar \imath j}e^{K/2}$ are finite this leads
to the conditions: 
\begin{equation}
W|_{\rm min}=0,\qquad \partial_A W|_{\rm min}=0, 
\qquad \partial_i W|_{\rm min}=0.
\label{partialb}
\end{equation}
If these conditions are satisfied, ${\cal N}=2$ supersymmetry is either 
partially broken to ${\cal N}=1$ or unbroken, depending on the
eigenvalues of the gravitino mass matrix eq.(\ref{gravmatb}).
Specifically, if the $e_I$ and $m^I$
 are real at a ${\cal N}=1$ minimum, or if
$e^{K/2}|_{\rm min}=0$, then both mass eigenvalues
in eq.(\ref{gravmatb}) are zero, and the full ${\cal N}=2$ supersymmetry
will be unbroken.

On the other hand, partial supersymmetry breaking to ${\cal N}=1$
supersymmetry is possible if some of the $e_I$ or $m^I$
are complex. In addition, according to \cite{FGP},
the existence of minima with partial supersymmetry breaking requires
that there exists a symplectic
basis in which the  periods $\tilde X^I$ are algebraically dependent.

\subsection{The superpotential from type IIB 3-form Fluxes}

\subsubsection{Calabi-Yau compactification}

In type IIB compactifications on a Calabi-Yau threefold  $M$ the superpotential
eq.(\ref{supo}) arises from turning on flux for NS and R three-form
field strength $H_{NS}^{(3)}$ and $H_R^{(3)}$ \cite{TayVa}.
The low energy spectrum consists, 
in addition to the ${\cal N}=2$ 
gravitational multiplet with the graviphoton,
of $N_V=h^{2,1}$ vector multiplets and $N_H=h^{1,1}+1$ hypermultiplets.
Turning on the internal H-flux manifests itself in the 4-dimensional effective
Lagrangian as a superpotential of the form
\begin{equation}
W=\int\Omega\wedge(\tau H_{NS}^{(3)}+H_R^{(3)}),\label{supoa}
\end{equation}
where $\Omega$ is the holomorphic 3-form on the Calabi-Yau space and $\tau$ the
complex type IIB couplings constant.
This superpotential is closely related to
NS5 resp. D5 branes  wrapped around 3-cycles ${\cal C}^{(3)}$
in the Calabi-Yau space. In the four-dimensional
effective theory the wrapped 5-branes correspond
to domain walls whose BPS tension
is  the jump $\Delta W$ of the  superpotential across the wall.

In order to bring eq.(\ref{supoa}) to the form (\ref{supo})
one expands the 3-cycles dual to the H-fluxes in terms of the
basis vectors $(A^I,B_I)$ of $H_3(M,{\bf Z})$ as
\begin{equation}
\tau {\cal C}_{NS}^{(3)}+{\cal C}_R^{(3)}=e_I(\tau)A^I-m^I(\tau)B_I,
\end{equation}
where $e_I(\tau)$ and $m^I(\tau)$ are
defined as 
\begin{equation}
e_I(\tau)=e^1_I\tau+e^2_I,\qquad m^I(\tau)=m^{1I}\tau+m^{2I}.\label{nm}
\end{equation}
The integer symplectic vectors
$(e^1_I,m^{1I})$ and $(e^2_I,m^{2I})$ are the quantized flux values
of the NS resp. R 3-form fields through the 3-cycles.
Then the superpotential (\ref{supoa}) becomes \cite{TayVa}
\beqa
W=\int_{{\cal C}_R^{(3)}}\Omega+\tau\int_{{\cal C}_{NS}^{(3)}}\Omega
& = &W_R+\tau W_{NS}=e_I(\tau)X^I-m^I(\tau)F_I,\nonumber \\
W_R & = & e_I^2X^I-m^{2I}F_I,\nonumber\\
W_{NS} & = & e_I^1X^I -m^{1I}F_I\ ;
\label{supoaa}
\eeqa
here $X^I=\int_{A^I}\Omega$ and $F_I=\int_{B_I}\Omega$.
Similar superpotentials were already discussed in \cite{FKLZ}.
This superpotential $W$ is $Sp(2N_V+2,{\bf Z})$ invariant. 
Under the type IIB S-duality transformations $\tau\rightarrow
{a\tau+b\over c\tau+d}$ it transforms with modular weight
$-1$,
\begin{equation}
W\rightarrow {W\over c\tau+d},\label{tausutr}
\end{equation}
provided that $(H_{NS}^{(3)},H_R^{(3)})$, or equivalently 
$(e_I^1,e_I^2)$ and $(m^{1I},m^{2I})$
transform  as vectors under 
$SL(2,{\bf Z})$.

\subsubsection{The scalar potential and the question of partial
supersymmetry breaking}

Next we have to determine the K\"ahler potential and the
scalar potential, 
which receives contributions from the scalar fields in both, vector and hypermultiplets. 
On the hypermultiplet side we are dealing with the complex dilaton
field $\tau$ and $Y=({\rm vol(CY)})^{1/3}+i~a$,  
the volume of the Calabi-Yau space and its axionic partner $a$, 
plus possibly other complex fields $q_i$. 
We assume that the vacuum expectation
values of the $q_i$ are zero and thus we can neglect them
in the following discussion; we are mainly interested in the
contributions of $\tau$ and $Y$ to the K\"ahler potential. We work from now on 
in the weak coupling limit, i.e. $\tau\rightarrow\infty$ and in the 
large volume limit, i.e. $Y\rightarrow\infty$. 
In these two limits the K\"ahler potential is 
explicitly known :
\begin{equation}K=K_V+K_H,\qquad K_H=
-\log({1\over 2 i} (\tau-\bar \tau))-3\log (Y+\bar Y).
\label{khyper}\end{equation}
The function $G=K+\log |W|^2$ is invariant under
$SL(2,{\bf Z})_\tau$.

Since the superpotential (\ref{supoaa}) does not depend on the
field $Y$, the contribution of $Y$ to the scalar potential $v$ has 
precisely the effect to cancel the negative vacuum energy.
Specifically, the scalar potential now is
\begin{equation}
v=e^G\biggl( G_{A}G_{\bar A}G^{A\bar B}+G_{\tau}G_{\bar\tau}
G^{\tau\bar\tau}\biggr).\label{scalarpotaa} 
\end{equation}
Supersymmetric minima of $v$ require that the auxiliary fields $h$
Since the superpotential does not depend on $Y$ and on 
the other hypermultiplets
$q_i$, the conditions $D_YW=D_{q_i}W=0$ imply $W=0$, and therefore the 
conditions of unbroken local ${\cal N}=1$ 
supersymmetry turn into the conditions of
unbroken global ${\cal N}=1$ supersymmetry, which read:
\begin{equation}
{dW\over dz^A}=0,\quad {dW\over d\tau}=0, \quad W=0.
\end{equation}

Using the specific form eq.(\ref{supoaa}) of $W$ these conditions
turn into
\begin{equation}
W_{NS}=0,\quad W_{R}=0, \quad {dW_R\over dz^A}+\tau{dW_{NS}\over dz^A}=0.
\end{equation}

Let us now consider the question of partial supersymmetry breaking and
compute the gravitino mass matrix (\ref{gravmatb}) for the
H-flux induced superpotential (\ref{supoaa}). With (\ref{supoaa},\ref{khyper})
the gravitino mass  matrix  (\ref{gravmatbb}) becomes
\begin{eqnarray}
S_{AB}={i\over 2(\tau-\bar \tau)^{1/2}(Y+\bar Y)^{3/2}}e^{K_V/2}
\left(\matrix{
      -W +2i\, \Im \tau\, W_{NS}  &      0     \cr 
  0     &      W}
\right) .\label{gravmatbd}
\end{eqnarray}
We see that a priori partial supersymmetry breaking (one vanishing eigenvalue) 
is only possible  in the presence of NS fluxes and ${\rm Im}(\tau) \neq 0$.
However if we treat $\tau$ as a dynamical field we
have to require that $dW/d\tau =0$ and hence $W_{NS}=0$. Therefore, as
soon as we are 
searching for ${\cal N}=1$ supersymmetric vacua, with $dW/d\tau =0$, 
both gravitino eigenvalues are zero and the theory is ${\cal N}=2$ 
supersymmetric. In other words, partial supersymmetry breaking 
seems to be impossible in connection with the H-flux induced superpotential 
eq.(\ref{supoaa});
supersymmetry is either
completely broken, or supersymmetric minima always preserve full
${\cal N}=2$ supersymmetry.

\subsubsection{The type IIB superpotential from gauged ${\cal N}=2$
supergravity}

Comparing the superpotential eq.(\ref{supoaa}) with the
general expressions in the previous section, we can derive the
following electric and magnetic Killing prepotentials
for the hypermultiplet fields ($\tau=\tau_1+i\tau_2$):
\beqa
P^1_I&=& e^{K_H/2}~\Im ~e_I={\tau_2^{1/2}\over (Y+\bar Y)^{3/2}}~ e^1_I, \nonumber\\ 
P^2_I&=&e^{K_H/2}~\Re ~ e_I={1\over\tau_2^{1/2} (Y+\bar Y)^{3/2} }(\tau_1 e^1_I-e^2_I),\nonumber\\
\tilde P^{1I}&=& e^{K_H/2}~\Im ~m^I={\tau_2^{1/2}\over(Y+\bar Y)^{3/2}}~ m^{1I}, \nonumber\\
\tilde P^{2I}&=&e^{K_H/2}~\Re ~ m^I={1\over\tau_2^{1/2} (Y+\bar Y)^{3/2}}(\tau_1 m^{1I}-m^{2I}).
\label{kilpre}
\eeqa
Similar as in \cite{Mich} these prepotentials should be obtained 
from the electric resp. magnetic gauging of the  hypermultiplets $Y$ and
$q=(S,C_0)$, where the complex dilaton field $S$ is the NS component of $q$,
and $C_0$ its complex Ramond component. Gauging in an $SL(2,{\bf Z})$ 
invariant way two particular isometries of the hypermultiplet moduli space 
${\cal M}_H=SU(2,1)/SU(2)\times U(1)$ will lead to the Killing prepotential
(\ref{kilpre})\footnote{We acknowledge discussions
with G. Dall'Agata and J. Louis.}.
Note that the condition that the electric and magnetic charges 
are mutually local is equivalent to the locality
of the Ramond and NS flux vectors, i.e. 
\begin{equation}
\int H_{NS}^{(3)} \wedge H_R^{(3)} \, \sim \, m \, \times \, e \, = \,
m^{1I}e^2_I-m^{2I}e^1_I \, = \, 0.
\end{equation}
As we will discuss in the last section, some special H-fluxes which do
not satisfy this constraint can also lead to supersymmetric vacua.

\subsection{The superpotential from type IIA fluxes}

Consider now type IIA compactification on the mirror Calabi-Yau space $W$
with Hodge numbers $h^{2,1}(W)=h^{1,1}(M)$ and $h^{1,1}(W)=h^{2,1}(M)$.
The number of vectormultiplets is $N_V=h^{1,1}(W)$, and the number of hypermultiplets is 
$N_H=h^{2,1}(W)+1$.
The type IIA superpotential can be obtained performing
the mirror map on the type IIA superpotential eq.(\ref{supoaa}).
Since the IIA mirror configuration to the wrapped IIB NS 5-branes is unknown, 
we discuss the case of turning on  Ramond
fluxes only. The mirror flux of $H_{R}^{(3)}$ corresponds to fluxes
of the IIA Ramond fields $H_R^{(6)}$, $H_R^{(4)}$ and $H_R^{(2)}$
which are dual to 0,2 and 4-cycles on $W$, plus one other flux 
term corresponding to the 6-cycle, $W$ itself.

We will define the IIA flux vectors with respect
to the integral basis $\Xi_\infty$ (see sect. 3.1). 
Then the IIA superpotential is \cite{TayVa}
\begin{eqnarray}
W&=&\int_W (H_R^{(6)}+J\wedge H_R^{(4)}+J\wedge J\wedge H_R^{(2)}+m^0~
J\wedge J\wedge
J)=\nonumber \\
&=&e_0\, +\, 
\int_{{\cal C}_R^{(2)}}J\, +\, \int_{{\cal C}_R^{(4)}}J\wedge J\, +\, m^0
\int_W J\wedge J\wedge J, 
\label{supob}
\end{eqnarray}
where $J$ is the K\"ahler class of $W$.
The corresponding domain walls are due to D2-branes, 
living in the uncompactified space, 
D4-branes wrapped around the 2-cycles ${\cal C}_R^{(2)}$,
D6-branes wrapped around ${\cal C}_R^{(4)}$ and D8-branes wrapped around
the entire CY-space $W$.
Next we choose a basis $J^A$ ($A=1,\dots
h^{1,1}$) for $H_2(W,{\bf Z})$,
\begin{equation}
{\cal C}_R^{(2)}=e_A J^A,
\end{equation}
as well as a dual basis $\tilde J_A$ for $H_4(W,{\bf Z})$
($J_A\wedge\tilde J_A=\Omega\wedge\bar\Omega$, no sum on $A$)
\begin{equation}
{\cal C}_R^{(4)}=m^A\tilde J_A.
\end{equation}
The integers $e_A$ and $m^A$ are the quantized fluxes of $H^{(4)}_R$ and
$H^{(2)}_R$ through the 4- and 2-cycles, respectively.
Then the type IIA  superpotential (\ref{supob}) can be written 
in homogeneous coordinates as
\begin{equation}
W=e_IX^I-m^IF_I,\label{supobb}
\end{equation}
where we have replaced the $e_0$ by $e_0X^0$.
Therefore the classical IIA periods $X^0$, $X^A$ are associated with the
0- and 2-cycles of $W$, whereas the periods $F_A$ and $F_0$ correspond
to the 4- and 6-cycles. The integers $(e_I,m^I)$ transform
as a vector under $Sp(2 h^{1,1}(W)+2,{\bf Z})$.

\subsection{Relation between the
superpotential due to
Ramond fluxes and ${\cal N}=2$ black holes and attractor mechanism}

The above discussion of the superpotential due to
internal fluxes has a close relationship to
extremal black hole solutions in ${\cal N}=2$ supergravity,
which we will now exhibit. We will show that the supersymmetry condition
$h^{\bar A}=0$ (see eq.(\ref{auxfields})) is formally analogous 
to the attractor equations which determines the values of the scalar
fields at the horizon of ${\cal N}=2$ supersymmetric black holes.

Consider ${\cal N}=2$ BPS states, whose masses are equal
to the central charge $Z$ of the ${\cal N}=2$ supersymmetry algebra.
The magnetic/electric charge vector of the BPS states is defined as 
\beqa
p^I&=&\frac{1}{2\pi}\int_{S^2}F^I,\nonumber\\
q_J&=&\frac{1}{2\pi}\int_{S^2}G_J,\label{bhcharges}
\eeqa
where $F^I$ and $G_ J$ are the electric and 
magnetic Abelian field strengths in four dimensions.
In terms of the charge vector $Q=(p^I,q_I)$ 
and the period vector $V$
the BPS masses are \cite{Ceresole}:
\beqa
M_{BPS}^2=|Z|^2=|\langle Q,V\rangle |^2=e^{K_V}|q_IX^I(z)-p^IF_I(z)|^2
\equiv e^{K_V}\,|{\cal M}(z)|^2.\label{bpsmasses}
\eeqa

Extremal ${\cal N}=2$ black holes solution leave half of 
the supersymmetries unbroken. They are BPS states. 
In type II string theory they can be constructed as D-branes wrapped around
the internal CY cycles, where the wrapping numbers corresponds to the electric 
and magnetic charges.
Specifically in type IIB, the black holes arise from wrapped D3-branes
around 3-cycles, whereas in type IIA black holes originate from
wrapping D6, D4, D2 and D0-branes over the cycles of the respective dimensions.

Near the horizon the values of the moduli fields, and thus the 
value of the central charge, 
are strongly restricted by the presence of full ${\cal N}=2$ supersymmetry.
In \cite{FerKal1} 
it was proved that this implies that the central charge becomes 
extremal on the horizon. As a consequence,  
independent of their asymptotic values, at the horizon the moduli 
are uniquely determined in terms of the  magnetic/electric charges $p^I$ and $q_I$. 
This is called the attractor mechanism. The value of the 
central charge at the horizon is related to the Hawking-Bekenstein entropy via
\be
{{\cal S}\over \pi} =  \vert Z_{\rm hor}\vert^2\;. \label{entropy}
\eq
In order to obtain the attractor values of the moduli at the 
horizon for extremal ${\cal N}=2$ black holes, one has to determine the extremal 
value of the central charge in moduli space. This implies 
\begin{equation}
\partial_A |Z |= 0 \quad\leftrightarrow\quad D_A{\cal M}=0\;.\label{blackholes}
\end{equation}       
These equations are difficult to solve in general.
They are, however, equivalent to the following set of algebraic equations
\cite{FerKal1} 
\beqa
\bar Z \, V - Z\,\bar V = i Q\, .\label{simplcond}
\eeqa
Several solutions of these equations in the context of
Calabi-Yau black holes were discussed in \cite{CLM,BCDKLM}.

Comparing the extremal black holes with the
${\cal N}=1$ supergravity action we  get the following 
formal correspondence between the BPS masses of the
${\cal N}=2$ black holes and the ${\cal N}=1$ superpotential,
\begin{equation}
{\cal M}\cong W, \quad {\rm with}~q\cong e,\quad p\cong m\, ,
\end{equation}
as well as the correspondence between the black hole entropy and
the gravitino masses,
\begin{equation}
{{\cal S}\over \pi}e^{K_H}\cong m_{3/2}^2.
\end{equation}
The
extremization of the central charges at the horizon, eqs.(\ref{blackholes})
and (\ref{simplcond}),
corresponds to the condition of vanishing auxiliary fields $D_AW=0$, 
i.e. to unbroken ${\cal N}=1$ supersymmetry.
The condition $m_{3/2}^2=0$ is  equivalent
to dealing with an extremal black hole with vanishing entropy.
Therefore the supersymmetric points of the effective supergravity action
precisely correspond to the attractor points of the ${\cal N}=2$
black holes.
These observations will turn out to be useful to find 
explicitly the points of
preserved ${\cal N}=1$ supersymmetry, since the equations $D_AW=0$ can be
translated
into the following equation for the symplectic vectors $V$ and $H$:
\begin{equation}e^{K_V/2}(
\bar W \, V-W \, \bar V )=iH \, . \label{susyeq}
\end{equation}

\section{Type II Vacua with  Ramond Fluxes}

\setcounter{equation}{0}

In this section we will consider type IIB compactifications with all
NS 3-form fields turned off. Then the  superpotential (\ref{supom}) 
does not depend on the scalar fields of the universal hypermultiplet. 
It also means that the fluxes are mutually local.
The condition of having unbroken supersymmetry
at the minimum of the scalar potential then has solutions only 
at subsets of the boundary of the moduli space \cite{TayVa,Mayr}.

\subsection{Type II compactifications on Calabi-Yau threefolds}

Let us review the aspects of the geometry which will be relevant for the
discussion of fluxes and the question of supersymmetry breaking. 
The vector moduli space is completely geometrical in the type IIB 
compactification on Calabi-Yau threefolds $M$.\footnote{For a review on string 
vacua with ${\cal N}=2$ supersymmetry see \cite{lust}} 
This special K\"ahler manifold  
arises as the moduli space of complex structure deformations of $M$ 
for the type IIB string compactification on $M$. By mirror symmetry, it 
is equivalent to the complexified K\"ahler structure deformation space on the 
mirror $W$ of $M$, which describes the type IIA string vector 
moduli space on $W$.

Let us now investigate the basis of the fluxes in type IIA/B
compactifications.
In type IIB compactifications on $M$ the fluxes of the $3-$form field strengths 
$H^{(3)}_{R}$ and $H^{(3)}_{NS}$ are w.r.t. an integral symplectic basis of 
$H^3(M)$. Following \cite{COGP,HKTY} we will find such an integral 
symplectic basis for the 
periods, or equivalently a basis for 
$H_3(M,{\bf Z})$ at the point of maximal unipotent monodromy, which corresponds 
in the mirror $W$ to the large volume limit $(R^A)^2\rightarrow \infty$.
At this point,  which is at $z^A=0$ in the coordinates used in \cite{HKTY}, 
one has a unique analytic period, normalized to $X^0=1+{\cal O}(z)$, 
and $m={\dim} (H^1(M,\Theta)) = h^{2,1}(M)$ logarithmic periods $X^A$, 
which provide natural special K\"ahler 
coordinates $t^A={X^A\over X^0}=
{1\over 2 \pi i} \log(z^A)+\sigma_A$, where $\sigma_A= {\cal O}(z)$ and 
$t^A:=B^A+i (R^A)^2$.  

The prepotential $F$, which is  homogeneous of degree 
two in the periods $X^I$, is ($q_A=\exp(2 \pi i t^A)$) 
\begin{eqnarray}
{ F}
&=&-{C_{ABC} X^A X^B X^C \over 3! X^0}+ n_{AB} {X^A X^B \over 2}+ c_A X^A X^0-i{\chi 
\zeta(3)\over 2 \pid}(X^0)^2+ (X^0)^2 f(q)\nonumber\\
&=& (X^0)^2{\cal F}= X_0^2\left[-{C_{ABC} t^A t^B t^C\over 3!}
+n_{AB} {t^A t^B \over 2}+
c_A t^A-i{\chi \zeta(3)\over 2 \pid} +f(q)\right]\ . \label{geomprep}
\end{eqnarray}
It defines an integral basis 
for the periods in the following way
(note that in the following the periods
$F_I$ are ordered in a different way compared to
eq.(\ref{sympsectionhol}))\footnote{This basis is unique up to 
integral symplectic transformations. E.g. a slightly more complicated choice 
has been made in \cite{COGP},  which amounts to a 
shift of the $C_A$ by an integer. 
Odd $C_{ABC}$  requires that some of the $n_{AB} \in {\bf Z}\setminus \{ 0 \}$ 
to get an integer monodromy around $z_A=0$. Using mirror symmetry and the 
expression for the $D4$-brane charge  one can determine $n_{AB}$ as the integral 
of $J_A\wedge J_B$ against $i_* c_1(D)$ 
($i:\, D\hookrightarrow W$) where $D$ is the divisor dual to 
$J_A\wedge J_B$ \cite{DR,hosono,mayrII}. }
\beq
\Xi_\infty=\left(\matrix{
X^0  \cr 
X^A \cr
{\p { F}\over\p X^A}\cr
{\p { F}\over\p X^0 }}\right)=X_0\left(\matrix{
1   \cr
t^A \cr
{\p {\cal F}\over \p t^A}\cr
2 {\cal F}- t^A  \p_A {\cal F } }\right)=X^0\left(\matrix{
1\cr 
t^A\cr
-{C_{ABC}\over 2} t^B t^C+{n_{AB}} t^B+c_A+\p_A f(q)\cr
{C_{ABC}\over 3!} t^A t^B t^C+c_A t^A-i{\chi \zeta(3)\over \pid}
+{\cal O}(q)}\right)\ .
\label{basis} 
\eeq
In the type IIA interpretation $C_{ABC}=\int_W J_A J_B J_C\ge 0$ 
are the classical intersection numbers, where $J_A$ are 
$(1,1)$-forms in $H^2(W,{\bf Z})$, which span the K\"ahlercone, 
$c_A={1\over 24}\int_W c_2 J_A$\footnote{Note that, up to the $n_{AB}$,  
the data needed to specify (\ref{basis}) are
those which give the topological classification of the three-fold $W$
as it follows from a theorem of C.T.C. Wall.}. 
In type IIA the $q$  expansion of $F$ around the large volume 
is a  world sheet instanton expansion.  
The explicit form $f(q)$ can be determined by mirror symmetry using 
the type IIB compactification on $M$.

Note that the point $q_A=0$, $\forall\ A$ corresponds, by mirror symmetry, 
to a very singular configuration of $M$ (it degenerates to intersecting 
hyperplanes), i.e. from the Type IIB perspective the 
large volume limit of $W$ corresponds to a 
complex structure degeneration of $M$, 
where partial susy breaking might occur. Away from this point, in a generic 
direction  in the complex structure moduli space, 
$M$ is regular and we do not expect any unbroken supersymmetry. 
Going away from the supersymmetric groundstate world-sheet instantons on 
the mirror $W$ will break supersymmetry \cite{Mayr}, but the dynamics 
generically drives the theory back to its supersymmetric vacuum.   

 Interesting effects may also 
occur at other singular points on the moduli space of $M$, like the 
conifold points. Here one particular IIB 3-cycle $A^1$ shrinks to zero size, 
i.e. $X^1_{\rm IIB}\rightarrow 0$. However that does not correspond 
to a shrinking  type IIA 2-cycle. Rather the whole quantum volume of $W$, 
i.e. the period $F_0$, vanishes 
\cite{PolStro}. In the next section we will discuss this and other 
degeneration points in the Calabi-Yau moduli space.

In the type IIA interpretation of the period vector $\Xi$, 
$t^A$  scales as the third root of the volume of the threefold $W$. 
This relates via (\ref{bpsmasses}) the first entry of (\ref{basis}) 
to the $D0$-mass,
the next $m$ entries to the BPS masses of the wrapped $D2$-branes, 
followed by the masses of the $m$ $D4$ and  finally the last entry  
to the $D6$-brane wrapped  around the whole Calabi-Yau manifold 
$W$\footnote{The moduli space of the 
$D0$-brane,  $W$ itself, has been identified with the moduli 
space of the $D3$-brane 
on the special Lagrangian torus (in $M$) fibered over the $S^3$, 
which vanishes at the generic 
conifold \cite{SYZ}.}. This identification of the basis of 
$H^3(M,{\bf Z})$ and $\bigoplus_{i=0}^3  H^{i,i}(W,{\bf Z})$ 
maps a IIB RR $3-$form $H^{(3)}_{R}$ to a linear combination in 
$\oplus_{i=0}^3 H^{i,i}_{R}$ of type IIA  RR forms.

\subsection{Points in the moduli space corresponding to a nonsingular CY}

In the absence of $H^{NS}$, all cycles can be chosen to be 
$A$-cycles, say $A_1$ and $A_2$. It is assumed that the periods 
$X^I=\int_{A_I} \Omega$ can serve as homogeneous coordinates 
in the moduli space, in particular that they are algebraically independent, 
${\partial X^I\over \partial X^J}=\delta^I_J$. 
If $e^{K} G^{A \bar B}$ (cf. eq.(\ref{auxfields}))
is finite at the point in the moduli space under consideration 
then the condition for unbroken supersymmetry is $W=0$ and 
$D_A W=(\partial_A+K_A)W=0$ $\forall A$. This is equivalent to $W={\rm d} W=0$. 
Here the derivatives are w.r.t. the inhomogeneous coordinates  
$t^A={X^A\over X^0}$. For a superpotential of the form (\ref{supoaa})
this is equivalent to 
${\rm d} W={\partial W\over  \partial X^I} {\rm d} X^I=0$ 
and requires that $H^{R}=0$.

When could $X$ fail to be a suitable parametrization for the complex 
structure moduli space? In fact this happens even at points 
parametrizing non-singular Calabi-Yau manifolds. Let us consider for 
example the mirror of the sextic 
$W=2 x_0^3+\sum_{i=1}^4 x_i^6-\psi\prod_{i=0}^4 x_i=0$ in ${\bf P}(2,1,1,1,1)$ 
discussed in \cite{kt,moore}. 
With the data $\chi=-204$, $C_{AAA}=3$ and $\int c_2 J=42$ we find 
an integral basis $\Xi_\infty$ at the point of maximal 
unipotent monodromy $z={1\over (6 \psi)^6}=0$ 
{}from (\ref{geomprep},\ref{basis}). 
This can be analytically continued to $\psi=0$. Here we find a basis of
solutions $\Xi_0=(w_2,w_1,w_0,w_5)$  with 
\beq
w_0=-{i\pi^4\over 3^5}\sum_{n=1}^\infty{e^{5n i\pi\over6}\over\sin{\pi n\over6}
\Gamma(n)\Gamma^4\!\left(1-{n\over3}\right)\!\Gamma^4\left(1-{n\over6}\right)}
\left({6 \psi\over 2^{1\over 3}}\right)^n, 
\qquad w_k=w_0(e^{2\pi i k\over 6} \psi)\ .
\eeq
The transformation  matrix  $\Xi_\infty=N\Xi_0$ is 
\beq
N=\left(\matrix{
-1&-1&\phantom{-}1&\phantom{-}1\cr 
\phantom{-}0&\phantom{-}0&-3&\phantom{-}0\cr
\phantom{-}0&-3&\phantom{-}3&\phantom{-}0\cr
\phantom{-}3&\phantom{-}0 &-9&-6\cr}\right)\ . 
\eeq
It follows from this that 
$X=6 X_1+3 F_1-9 X_0-2 F_0=c\psi^2+{\cal O}(\psi^3)$
\footnote{
$c={i \pi^4 2^{10/3}\over 9 \sqrt{3} \Gamma\left({1\over 3}\right) 
\Gamma\left({2\over 3}\right)}$} 
is not a good coordinate for the moduli space and with $W=X$, 
$W={\rm d} W =0$ can be fulfilled.  
However, we see that due to the degeneration of the factor 
$e^K\sim {1\over |\psi|^2}$ (the metric stays finite) 
the  scalar potential does not vanish\footnote{$\tilde c=
{8\pi^2\over 177147\sqrt{3}\Gamma^2\!\left(1\over 3\right)
\Gamma^6\!\left(2\over 3\right) \Gamma^{16}\!\left(5\over 6\right)}$} 
\beq
V=e^K G^{\psi\bar\psi}|D_\psi W|^2
=\tilde c+{\cal O}(|\psi|^2) \ , 
\eeq
and supersymmetry is broken.

The periods at a generic point $\psi_0$ of the moduli space are all power series in the deformation 
parameter $\tilde a=\psi-\psi_0$ of the Calabi-Yau space.  
For simplicity consider a one 
parameter family $h^{11}=1$ and with Ramond fluxes only. We may choose 
a new variable $a$ as a fractional power of $\tilde a$ so that only
integer powers of $a$ appear in the periods.  Let us be concrete and consider 
the generic expansion of the
periods around the point $a=0$
\beq
\Xi_k=\sum_{n=0}^\infty c_{k,n}a^n,
\eeq
where generically all coefficients $c_{k,n}$ are non-zero. Iff at the point 
$a=0$ the first two coefficients 
$c_{k,n}$, $n=0,1$, of the periods would be linear dependent
over the rational numbers, then we could pick a
flux whose dual 
$X^A$ is not a good variable of the moduli space at $a=0$,
i.e. $X\sim a^2+{\cal O}(a^3)$. 
More precisely, if $X=x_k \Xi_k$, the $x_k$ have to satisfy  
\beq 
\sum_{k=1}^{2 h_{11}+2} x_k c_{k,n}=0, \qquad n=0,1 
\label{susycond}
\eeq
For the statement that all choices of $A$-cycles lead to good 
algebraically independent 
coordinates to hold\footnote{The 
infinitesimal Torelli Theorem implies only that there is {\it one} choice of 
$h_{2,1}+1$ cycles, whose
periods  can serve  as good homogeneous parameters.} the 
ratios ${x_k\over x_l}$ should be irrational for all $k,l$.
Clearly the above equation can be solved for $x_k\in {\bf C}$. 
An interesting question is whether it can be
solved for $x_k\in {\bf R}$. In this case it would be 
possible to select fluxes with 
``large'' integer  coefficients which would lead with an 
arbitrary precision to a supersymmetric vacuum
at the point $a=0$. To check this we plotted 
\beq
\det\left(\matrix{
\Re(c_0^1) &\ldots \Re(c_0^{4})\cr 
\Im(c_0^1) &\ldots \Im(c_0^{4})\cr
\Re(c_1^1) &\ldots \Re(c_1^{4})\cr 
\Im(c_1^1) &\ldots \Im(c_1^{4})}\right)
\eeq       
for the quintic hypersurface in 
${\bf P}^4$ and found that is vanishes only at the orbifold point, 
which implies that there is not even approximate supersymmetry 
for any choice of RR fluxed for a generic quintic.

\subsection{Overview over the degenerate cases}

As already emphasized, supersymmetry will be broken at generic points
in the Calabi-Yau moduli space.
However there is in fact the chance that supersymmetric minima
exist at those points where the Calabi-Yau space degenerates.
These points  correspond to limits where certain cycles
of the Calabi-Yau space shrink to zero size (resp. grow to infinity).
As we will see these degenerate points will correspond to
supersymmetric vacua in case the flux vectors are 
precisely aligned along the directions of the vanishing cycles. 

Let us analyze the situation in more detail. Suppose that we are
considering a superpotential of the form $W=e_1X^1$, where e.g. in type IIA,
$X^1$ corresponds to a two-cycle, $X^1=\int_{{\cal C}^{(2)}_1}J$ and
the flux is due to
a non-vanishing four-form $H^{(4)}_R$: $e_1=\int_{{\cal C}^{(4)}_1}H_R^{(4)}$.
Then, at  a generic point in the moduli space, where $X^1\neq 0$, the condition
$W=dW=0$ implies that $H_R^{(4)}=0$, i.e. non-vanishing flux necessarily
breaks supersymmetry. On the other hand, in case the two-cycle
vanishes, $X^1\rightarrow 0$, the condition $W=0$ is automatically satisfied.
Moreover the metric factor $e^KG^{1\bar 1}$  vanishes in many examples
at the
points where $X^1=0$. Hence (\ref{auxfields}) is also  satisfied.
We will show in the following that the degeneration points also correspond
to minima of the scalar potential, which means that they are supersymmetric
ground states of the theory. As mentioned already, the values of the 
scalar fields at these points precisely agree with the attractor
points in the context of supersymmetric black holes. So it is
quite natural to assume that the compactification is dynamically driven
to the attractor points in case we turn on H-fluxes which are aligned
along vanishing cycles.

Before we proceed let us first give a brief overview over the degenerate
cases with vanishing cycles
in the IIA moduli space and the
corresponding Ramond fluxes which are turned on.
For simplicity consider for the moment
a model with $h^{1,1}=2$; the corresponding
two moduli are $S=-iX^1/X^0$ and $T=-iX^2/X^0$. In the large volume ${\rm Re} S, {\rm Re}T\gg 1$ 
and ${\rm Re}S>{\rm Re}T$ limit, where $F\sim i S T^2$,  we have the following correspondences 
(we assume that the CY is a K3 fibration over a ${\bf P}^1_b$ base;
the K3 fiber contains a second ${\bf P}^1$, denoted by ${\bf P}^1_f$): 
\begin{eqnarray}
X^0 \quad & & \Longleftrightarrow\quad {\rm vol}({\cal C}^{(0)}),\nonumber\\
X^1\sim i S 
\quad & & \Longleftrightarrow\quad {\rm vol}({\cal C}^{(2)}_1)
\sim {\rm vol}({\bf P}^1_b),\nonumber\\
X^2 \sim i T
\quad &  &\Longleftrightarrow\quad {\rm vol}({\cal C}^{(2)}_2)
\sim {\rm vol}({\bf P}^1_f),\nonumber\\
F_1 \sim i{\partial F\over\partial S}\sim T^2
\quad & & \Longleftrightarrow\quad {\rm vol}({\cal C}^{(4)}_1)
\sim {\rm vol}(K3),\nonumber\\
F_2 \sim i{\partial F\over\partial T}\sim  ST
\quad & & \Longleftrightarrow \quad {\rm vol}({\cal C}^{(4)}_2),\nonumber \\
F_0 \sim i S T^2\quad & & \Longleftrightarrow 
\quad {\rm vol}({\cal C}^{(6)})\sim {\rm vol}(CY).
\label{cyclesst}
\end{eqnarray}
Here ${\cal C}^{(d)}$ is a cycle of real dimension $d$. 
${\rm vol}({\cal C}^{(0)})$ is a constant and 
${\cal C}^{(4)}_2$ is a four-cycle which contains the base ${\bf P}^1_b$ 
and the ${\bf P}^1_f$. All volumes are meant to be complexified volumes.
In the following the aligned fluxes will correspond to the
vanishing cycles. However other, non-aligned 
choices are of course also possible and will be mentioned in the
sects. 3.4-3.9.

\vskip0.3cm
\noindent {\it (i) The perturbative heterotic limit}

This is simply the limit where, in the heterotic dual, 
we turn off all instanton effects,
i.e. $S\rightarrow\infty$. Therefore, comparing
with (\ref{cyclesst}) we see that the cycles 
${\cal C}_1^{(2)}$, ${\cal C}_2^{(4)}$, ${\cal C}^{(6)}$ 
become large, which can be alternatively interpreted to 
mean that the remaining cycles  ${\cal C}^{(0)}$, 
${\cal C}^{(2)}_2$ and ${\cal C}^{(4)}_1$ vanish.
Turning on the aligned fluxes  $e_0$, $e_2$ and $m^1$,   the
superpotential takes the form
\beqa
W=e_0X^0+e_2X^2+m^1F_1 \ .
\label{supals}
\eeqa
Note that the periods $X^2$ and $F_1$ (and $X^0=1$ anyway)
do not vanish.  The superpotential is zero at the minimum, and $\dd W=0$, 
nevertheless supersymmetry will be unbroken because of the $e^K$ factor, 
cf. the discussion  before (\ref{partialb}).

\vskip0.3cm
\noindent {\it (ii) The large volume limit}

In this limit all IIA K\"ahler moduli are large: $S\rightarrow\infty$,
$T\rightarrow\infty$. This can be interpreted as having $X^0$ as
vanishing cycle. Using the aligned flux $e_0$, one derives the following
superpotential 
\beqa
W=e_0X^0 \, .
\eeqa

\vskip0.3cm
\noindent {\it (iii) The conifold limit}

The conifold limit is the limit where
one of the IIB three-cycles ${\cal C}^{(3)}_{IIB}$ shrinks to zero size.
As already observed in \cite{PolStro}, this conifold limit corresponds
in IIA compactification to the limit where the entire quantum six-volume of
the CY vanishes, i.e.
\beqa
F_0\rightarrow 0.
\eeqa
Hence we turn on the aligned flux $m^0$, and the
superpotential takes the form
\beqa
W=m^0F_0.
\eeqa

\vskip0.3cm
\noindent {\it (iv) The field theory Seiberg-Witten limit}

In this limit non-perturbative monopoles or dyons become massless \cite{sw}.
In case of massless monopoles, $u\rightarrow 1$ corresponding to
$a_D\rightarrow 0$, in the notation of \cite{sw}.  
The string interpretation of this situation is
a double scaling limit, namely the intersection of the conifold
limit with the large $S$ limit, which can be regarded as going to 
the $u$-plane. Hence we expect  (actually more cycles 
vanish, see sect. 3.7)
\beq
F_0\rightarrow 0,\qquad {1\over 2}F_1+iX^2\rightarrow 0.
\eeq
(The first line in this equation describes the conifold limit, 
whereas the second
line corresponds to going to the $u$-plane divisor.)
The corresponding superpotential with aligned fluxes will turn out to be
\beqa
W=m^0F_0+
m(F_1+2 iX^2) .
\eeqa

Finally, vanishing dyon masses correspond to $u\rightarrow -1$,
$a_D-a\rightarrow 0$.
Since the sum of the dyon electric/magnetic
charges plus the monopole charges equals the $W^\pm$--boson
charges, we can conclude that
\beqa
F_0-X^0 +{1\over } F_1 \rightarrow 0,\quad  \quad {1\over 2}F_1+iX^2\rightarrow 0.
\eeqa
The superpotential is
\beq
W=-m^0 (2 F_0- 2 X^0+ F_1) + m^1( F_1+2 iX^2) \  .
\eeq

\vskip0.3cm
\noindent {\it (v) The strong coupling limit}

The strong coupling singularity \cite{strong}
is an example of a degeneration where two intersecting cycles shrink to zero size. 
In case of non-vanishing NS and Ramond fluxes this degeneration leads to 
a situation in which the fluxes are non-local w.r.t. each other.
To be specific consider a type IIA compactification on the 
Calabi-Yau manifold ${\bf P}^4(1,1,2,8,12||24)$ with $h^{1,1}=3$ and
$h^{2,1}=243$ which is an elliptic fibration over the Hirzebruch surface
${\bf F}_2$. The three vector moduli are $S$, the volume of the ${\bf P}^1$
basis of ${\bf F}_2$, $U$, the volume of the ${\bf P}^1$ fiber of ${\bf F}_2$,
and $T$, the volume of the elliptic fiber $E$. 
At the strong coupling point $S=0$ the following two cycles
with non-trivial intersection number shrink to zero size:
\beqa
& &{\rm vol}({\cal C}^{(2)}_S)\rightarrow 0\quad \Longleftrightarrow  \quad
S\rightarrow 0,\nonumber\\
& &{\rm vol}({\cal C}^{(4)}_S)- {1\over 2 }{\rm vol}({\cal C}^{(4)}_U)
\rightarrow 0\quad \Longleftrightarrow  \quad
F_S - {1\over 2} F_U \rightarrow 0,
\eeqa
where $C^{(2)}_i\cap C^{(2)}_j=\delta_{ij}$. 
At this point a $U(1)$
gauge group is enhanced to $SU(2)$, and also an $SU(2)$ adjoint
hypermultiplet becomes massless. Hence the corresponding (${\cal N}=2$)
$\beta$-function vanishes.
The superpotential with aligned fluxes is then
\beq
W=ie_SS+im(2 F_S-F_U)\, .
\eeq

\subsection{The IIA large volume limit and the perturbative heterotic limit}

\subsubsection{The classical heterotic limit}


In this section we consider the classical heterotic limit, or 
equivalently in IIA
language the limit where the base of the K3 fibration is large 
(see the appendix for more details),
i.e. $S\rightarrow\infty$. In this limit 
the prepotential is
\beqa
F= i(X^0)^2 S (T^a\eta_{ab}T^b+1),\label{hetprepot}
\eeqa
where $S=-iX^1/X^0,\,T^a=-iX^a/X^0$.
The corresponding period vector is then ($X^0=1$)
$(X^I,F_I)=(1,iS,iT^a;-iST^a\eta_{ab}T^b+iS,T^a\eta_{ab}T^b,2 S\eta_{ab}T^b)$.

As discussed before, we want to choose the H-fluxes aligned
with the directions of the vanishing cycles $X^0$, $X^a$ and $F_1$.
This leads to the following non-vanishing fluxes $e_0$, $e_a$, $m^S$, and
the superpotential takes the following form:
\begin{equation}
W=\tilde e_0-m^S T^a\eta_{ab}T^b+ie_aT^a\, ,\label{stualg}
\end{equation}
where $\tilde e_0=e_0-m^s$.
The ${\cal N}=2$
supersymmetric zero energy ground state, $W=0$ and $e^{K\over 2} DW=0$, is obtained for the
following (attractor) values of the (real) moduli:
\beqa
e_aT^a=0,\quad \tilde e_0-m^S T^a\eta_{ab}T^b=0, \quad S=\infty\, .\label{atpoint}
\eeqa
On the other hand, for finite $S$ supersymmetry is completely broken.

To be concrete, let us investigate in more detail the
corresponding scalar potential for the $STU$ model, 
assuming, for simplicity, that
the three moduli are real:
\beqa
v={\tilde e_0^2+(m^S)^2T^2U^2+e_T^2T^2+e_U^2U^2\over 2 STU}\label{potstu}
\eeqa
In the directions of $T$ and $U$ this scalar potential has its minima
at the ${\cal N}=2$ supersymmetry 
preserving points 
\begin{equation}
T_{\rm min}=-{e_U\over e_T}U_{\rm min},
\qquad U_{\rm min}=\pm\sqrt{-{\tilde e_0e_T\over m^Se_U}},
\end{equation}
where we need $-{e_T\tilde  e_0\over m^S e_U}>0$ for real moduli fields, 
as we assumed here. 
In the direction of the $S$-field the scalar potential has no minimum,
but has a run-away behavior, $v\sim 1/S$, which drives the $S$-field
to infinity. 

The {\it classical field theory limit} 
is naturally contained in this discussion. 
This is the limit, where we turn
off all field theory quantum effects, and
the perturbative $W^{\pm}$--bosons become massless.
In string theory, this limit
corresponds to a double scaling limit, namely to $S\rightarrow\infty$
together with $T\rightarrow U$.
Specifically, within the $STU$
model this limit is achieved by choosing $\tilde e_0=m^S=0$ and $e_T=-e_U$
in the superpotential eq.(\ref{stualg}), i.e. $W=ie_T(T-U)$.
The corresponding minima are at the line $T=U$, where the classical gauge
group is enhanced to $SU(2)$.

In summary, using the classical heterotic 
prepotential and turning on aligned fluxes one finds 
supersymmetric
minima with vanishing potential for finite $T_a$  and
infinite $S$.
In fact, for $S\rightarrow \infty$, 
the whole ${\cal N}=2$ supersymmetry is restored. 

\subsubsection{The large volume limit}

In this limit all K\"ahler moduli $t^A$ are large, $\Im(t^A)\rightarrow\infty$,
which means that all rational instantons are suppressed. The prepotential
is determined by the intersections numbers
$C_{ABC}$ and has the form
\begin{equation}
F^{\rm IIA} = -{1\over 6} C_{ABC}{X^AX^BX^C \over X^0},
\label{ftypea2a}
\end{equation}
where the IIA K\"ahler moduli are defined as $t^A=X^A/X^0$ ($\Im t >0$).
The K\"ahler potential is in this limit
\begin{equation}
K=-\log\Biggl( {i\over 6}C_{ABC}(t^A-\bar t^A)(t^B-\bar t^B)(t^C-\bar t^C)
\biggr) .
\end{equation}

Let us first discuss the case of aligned fluxes, i.e. $e_0$ is the only
non-vanishing flux, which leads to the superpotential 
$W=e_0X^0$. This case is essentially contained in the discussion
of the previous section (see eq.(\ref{potstu})). 
The scalar potential can be
computed in a straightforward manner (see also \cite{PolStro}):
\begin{equation}
v=4\, e_0^2\, e^K\sim {(e_0)^2\over {\rm vol}(CY)}\, .
\end{equation}
This potential has no extrema for finite values of $t^A$, but it shows 
the characteristic run-away behavior, which drives all moduli to infinity,
where supersymmetry is unbroken. 

As an alternative let us consider a choice of Ramond fluxes which are not
aligned with the vanishing cycle ${\cal C}^{(0)}$.
Specifically, we decide to turn on the flux $e_0$ and all fluxes $m^A$,
which correspond to the 4-cycles ${\cal C}^{(4)}_A$.
The superpotential is now
\begin{equation}
W=X^0(e_0+{1\over 2 }m^A C_{ABC}t^B t^C).\label{supoIIA}
\end{equation}
The conditions for  preserving supersymmetry
in the sector of the fields $t^A$, $h^A=0$, are then solved, 
using the attractor equations (\ref{susyeq}), by 
the following values of the moduli \cite{BCDKLM}:
\begin{equation}
t^A_{\rm Susy}=-im^A\sqrt{-{6e_0\over C_{ABC}m^Am^Bm^C}}.\label{solIIA}
\end{equation}
(We haven chosen $e_0<0$ and $m^A>0$.)
Consistency with the large volume limit requires $-e_0\gg m^A$.

For the gravitino mass $m_{3/2}^2=e^G$ at the points (\ref{solIIA}) we find
\begin{equation}
m_{3/2}^2|_{{\rm Susy}}=2\sqrt{-e_0{C_{ABC}\over 6}m^Am^Bm^C}.
\end{equation}
Note that this
expression is identical to the entropy ${{\cal S}\over\pi}$ 
of classical Calabi-Yau
black hole solutions which are due $m^A$ D4-branes, wrapped around
the CY 4-cycles, plus $e_0$ D0 branes.
Since $C_{ABC}m^Am^Bm^C \neq 0$, $m_{3/2}$ is non-zero.

Analyzing the scalar potential it turns out that
the points (\ref{solIIA}) are indeed extrema of $v$; 
at these extrema the potential has the value:
\begin{equation}
v|_{{\rm Susy}}=2\sqrt{-e_0{C_{ABC}\over 6}m^Am^Bm^C}.
\end{equation}
Although the auxiliary fields $h^A $ are zero at the extrema of $v$,
supersymmetry is nevertheless broken in the sector of
the hypermultiplets $\tau$ and $Y$. This comes from the fact that $W\neq 0$
at the points (\ref{solIIA}), and hence  $h^\tau,h^Y\neq 0$.

To be more specific about the nature of the extrema of $v$, let us
compute $v$ for the $STU$ model with $C_{STU}=1$ 
and all other $C_{ABC}=0$ for real moduli $S,T,U$ :
\beqa
v= {e_0^2+(m^1)^2 T^2 U^2+(m^2)^2 S^2U^2
+(m^3)^2 S^2 T^2\over 2 S\, T\, U}\, .
\eeqa
We see that 
this potential indeed possesses a minimum at
the points eq.(\ref{solIIA}). Therefore  the model
with non-aligned fluxes exhibits
a stable non-supersymmetric ground state with positive scalar
potential at its minimum.
On the other hand, since the moduli $t^A$ at the minimum of the potential
are large (for$-e_0 \gg m^A$) but not infinite, one should also discuss
the contribution of the exponentially suppressed instanton terms to
the prepotential, where we expect that 
the minima of the potential are shifted by corrections of the order
$e^{-t^A}$.

\subsection{The conifold locus}

In this section we want to explore the vacuum structure of
a type IIB compactification near the conifold locus in the moduli space.
The generic conifold locus is the  co-dimension one locus
in ${\cal M}$, where in $M$ a cycle, say  $A_1$,  with the topology of 
$S^3$ vanishes, while the remaining $3$-cycles stay finite.
More precisely the Calabi-Yau space $M$ exhibits a nodal singularity, i.e. 
it is described locally by the eq. $\sum_{i=1}^4 \epsilon_i^2=\mu$. For
$\mu\rightarrow 0$ the real part of this local 
equation describes the vanishing 
$S^3$. In the vicinity of a conifold point, 
$X^1=\int_{A_1} \Omega\rightarrow 0$, an 
additional hypermultiplet, the ground state of a singly wrapped $3-$brane 
around $A_1$, with mass proportional to $|X^1|$ becomes light 
\cite{stromon}. It is charged w.r.t. to the 
$U(1)^{N_{V}}$ gauge symmetry of the vector 
multiplets\footnote{E.g. it corresponds to a magnetic monopole or dyon in
the effective gauge theory, whereas its effective supergravity description
is given as a massless black hole.}. Related ${\cal N}=2$ black hole solutions 
at the conifold locus were considered before in \cite{stromon,BLS}.

In the following we will discuss the simplest situation with periods
$X^i=\int_{A_i} \Omega$, $i=0,1$ and the dual periods 
$F_i=\int_{B^i}\Omega$ (with $A_i \cap B^j=\delta_i^j$), 
where $X^1$ vanishes 
at the conifold locus and the other periods remain finite.
 So at the conifold we have
\be
\mu :={X^1\over X^0} = 0\ .
\eq
In addition, if the 3-fold is transported
along a closed loop around the conifold locus the period
$F_1$ undergoes a monodromy transformation, given by the Lefshetz formula 
\be
F_1 \rightarrow F_1 + X^1  \ ,
\eq
while all other periods have trivial monodromy. Therefore the 
periods near the conifold have the expansion\footnote{For the field 
theory interpretation it 
is essential that  $c^{(0)}_i\neq 0$, otherwise a magnetically as well as an 
electrically charged particle become massless at $\mu=0$. 
The ${\cal O}(\mu^n)$ parts are fixed by the Picard-Fuchs equation.}
$F_1=\sum_{i=0}^2c^{(i)}_1\mu^i+{X^1(\mu)\over 2 \pi i}\log(\mu)
+{\cal O}(\mu^3)$, $X^1=c\mu+{\cal O}(\mu^2)$, 
$F_0=\sum_{i=0}^2c^{(i)}_0\mu^i +{\cal O}(\mu^3)$ and $X^0
=\sum_{i=0}^2 c^0_{(i)}\mu^i+{\cal O}(\mu^3)$ and near $\mu=0$ 
the prepotential can be expanded as
\be
F = -i \, (X^0)^2 \left(a + {c^2 \over 4 \pi}  \mu^2 
\log \mu+ b \mu\,  + (\mbox{analytic terms})  \right) \ .
\eq
It is easy to see that the K\"ahler potential is finite
at the conifold point:
\be
e^{-K_V}=4\, {\rm Im}\, a.
\eq
In contrast, the internal moduli space metric logarithmically diverges at the 
conifold point:
\be
K_{\mu\bar \mu}=-\log |\mu| {1\over 2 {\rm Im} a}.
\eq

In the type IIA mirror compactification on the mirror quintic $W$ the 
conifold point for $M$ 
corresponds to the limit where the entire quantum volume 
of $W$  shrinks to zero size, whereas the other cycles stay 
finite \cite{COGP}.  $\chi=-200$, $\int c_2 J=50$ and $n_{11}=1$ \cite{COGP} fixes 
the integral basis (\ref{basis}). Using the  relation $\mu=1-5^5 z$, where 
$z$ is the variable at the large complex structure point $z=0$ \cite{HKTY}, 
$c$ turns out to be ${\sqrt{5}\over 2 \pi i}$ and one can 
fix the $c_I^{(j)}$, $j=1,2$  
so that after analytic continuation\footnote{To six significant 
digits, they are 
$c^{(0)}_1=1.07073$, 
$c^{(1)}_1=-.0247076$,  
$c^{(2)}_1=.0566403$, 
$c^{(0)}_0=6.79502- 7.11466 i$, 
$c^{(1)}_0= 1.01660-.829217 i$,  
$c^{(2)}_0=.711623-.580451 i$, 
$c_{(0)}^0=1.29357 i$, 
$c_{(1)}^0=-.150767 i$ and  
$c_{(2)}^0=.777445 i$.} $X^{\infty,0}=-F_1$, $X^{\infty,1}=-F_0$, $F^{\infty}_1=X^0$ and 
$F^{\infty}_0=X^1$. One checks that $X^I,F_I$ 
satisfy the integrability conditions for the existence of the prepotential 
and determines the constants in $F$: $a=0.517061+i .04500226$ and $b=0$. 
Even for this choice
of coordinates in one parameter models the constant $a$ is not universal, 
as we find 
for the sextic in ${\bf P}(1,1,1,1,2)$ $a=.147507\,i$, but $b=0$ for all one moduli cases.

For hypersurfaces in toric varieties with an arbitrary number of 
moduli we find that the $S^3$ vanishing at the principal discriminant 
\cite{HKTY} corresponds via mirror symmetry always to the quantum 
volume $F^{\infty}_0$.

Consider the case where the flux that is aligned with the vanishing
cycle of the conifold point is turned on. In type IIB the corresponding
superpotential is
\begin{equation}
W=e_1 \mu .
\end{equation}
Since $W=0$ but $\partial W/\partial \mu =e_1\neq 0$  at the conifold point, one
might expect that the conifold point does not correspond to a supersymmetric
ground state with vanishing scalar potential. However this conclusion
is not correct in the context of supergravity, since the K\"ahler
metric diverges at the conifold point. In fact, the corresponding
supergravity scalar potential in the vicinity of the conifold point,
\begin{equation}
v=|W_\mu|^2e^K K^{-1}_{\mu\bar \mu}=-{e_1^2\over \log |\mu|^2},
\end{equation}
has a supersymmetry preserving minimum at $\mu=0$ with $v=0$.
Hence $\mu$ is attracted to the conifold point \cite{PolStro}. 

This ground state of supergravity is not changed if  also the additional light
hypermultiplet is 
included into the superpotential at the conifold point \cite{TayVa}:
\begin{equation}
W=e_1\mu \, +\, \mu \phi\tilde\phi \, .
\end{equation}
Now the supersymmetric, stationary points of $v$ are at
$W=0$ and $dW=0$, which leads again to $\mu=0$ and in addition to
$\phi\tilde\phi=-e_1$, as discussed in \cite{TayVa}.

\subsection{Colliding conifold loci}

Next we study the situation where two conifold loci meet. (This corresponds
to an $A_2$ singularity, cf. below. The generalization to
$A_n$ singularities is straightforward.)
To be concrete we consider the mirror of the $X_{18}(1,1,1,6,9)$ Calabi-Yau hypersurface, a two parameter model 
where two conifold loci meet \cite{Candelas:1994hw}. Many aspects of the generalization to the meeting 
of several conifold loci are  straightforward.  $X_{18}(1,1,1,6,9)$ is an elliptic fibration
over ${\bf P}^2$ and  the mirror manifold may be defined by
\beq
P=x_1^{18}+ x_2^{18}+x_3^{18}+x_4^3+x_5^2 -
3 \Phi x_2^6 x_2^6x_3^6-6^{2\over3} \Psi x_1x_2x_3x_4x_5=0 \ ,
\eeq
with the orbifold action $Z_{18}\times Z_{18}$: 
$(x_1\mapsto x_1 \exp {2\pi i \over 18},x_2\mapsto 
x_2 \exp(17{ 2\pi i \over 18}))$ 
$(x_1\mapsto x_1 \exp {2\pi i \over 18},x_3\mapsto 
x_3 \exp(17{ 2\pi i \over 18}))$ 
on the coordinates. The canonical large complex structure 
coordinates are \cite{HKTY}
\begin{eqnarray}
z_1&=&-{\Phi\over \Psi^6}, \qquad l_1=(0,0,0,2,3,1,-6),\nonumber\\
z_2&=&-{1\over \Phi^3}, \qquad l_2=(1,1,1,0,0,-3,0)\ . 
\end{eqnarray}

Here the $l_i$ are the generators of the Mori cone, which identify the complex
coordinates of the mirror near $z_i=0$ as ${\log(z_1)\over 2\pi i}\sim t_E$, 
${\log(z_2)\over 2\pi i}\sim t_{{\bf P}^1}$, with the K\"ahler parameter $t_E$ of $X_{18}(1,1,1,6,9)$ 
measuring the size of the elliptic fiber and $t_{{\bf P}^1}$ measuring the 
size a ${\bf P}^1$ 
in the base ${\bf P}^2$. We find the discriminant by solving ${\partial P\over 
\partial x_i}=P=0$ 
for $\Phi$, $\Psi$ and $x_i$. The principal discriminant is the 
solution where all $x_i\neq 0$. We find $x_5=3 (\Psi x_1x_2x_3)^3$, 
$x_4=6^{1\over 3}(\Psi x_1x_2x_3)^2$, $x_1^{18}=x_2^{18}=x_3^{18}\neq 0 $ and 
\beq
\Delta_{con_1}=1-(\Psi^6+\Phi)^3=0\ .
\eeq 
A second solution is $x_4=x_5=0$, $x_1^{18}=x_2^{18}=x_3^{18}\neq 0$ and
\beq
\Delta_{con_2}=1-\Phi^3=0 \ .
\eeq  
Near $\Psi=0$ and $\Phi=1$, where $\Delta_{con_1}=\Delta_{con_2}=0$, 
the local expansion of the manifold is 
\beq
\epsilon_1^2+\epsilon_2^2+\epsilon_3^2+\epsilon_4^3=a \epsilon_4 +b\ ,
\eeq
the three-dimensional version of an $A_2$ singularity. In particular we have  
two $S_1^3$  and $S_2^3$ with $S_1^3 \cap S_2^3=1$, but as three is odd we 
have $S_i^3 \cap S_i^3=0$.

In the $z$ coordinates we have, up to irrelevant factors, 
$\Delta_{con_1}=1-3 z_1+3 z_1^2 -(1+z_2) z_1^3$ and 
$\Delta_{con_2}=1+z_2$, so the conifolds collide at the two points 
$(z^{\pm}_1={1\over 2} \pm {i\over 6} \sqrt{3},z_2=-1)$.
At these points we can solve the Picard-Fuchs equation in the variables  
$x_1=(1-{z_1\over z^{\pm}})$ and $x_2={1+z_2 \over 1-{z_1\over z^{\pm}}}$.
We get, as expected, two unique vanishing solutions $X^{cc,1}=x_1
+{\cal O}(x^2)=\int_{S^3_1}\Omega$ and 
$X^{cc,2}=x_1 x_2+{\cal O}((x_1 x_2)^2)=\int_{S^3_2}\Omega$ with dual periods 
$F^{cc}_1={1\over 2 \pi i} X^{cc,1} \log(X^{cc,1})+a_1 +holom.= \int_{T^3_1}\Omega$ and 
$F^{cc}_2={1\over 2 \pi i} X^{cc,2} \log(X^{cc,2})+a_2 +holom.= \int_{T^3_2}\Omega$. 
Further we can see, in this case by analytic continuation, 
that $T_1^3 \cap T_2^3=0$ and furthermore that the remaining two  periods 
$X^{cc,0},F^{cc}_0$ start with a constant term. 
This implies that to leading  order the Weil-Petersson-metric near the conifold is  
\beq
G_{x_i,\bar x_j}=-\left(\matrix{c_1\log(|x_1|) & 0 \cr
0  &c_2 \log(|x_2|)}\right),
\eeq
with $c_1,c_2 > 0$. 
The relation between $X^{cc}_i$ and the periods at infinity can be 
obtained via analytic continuation.
One finds $X^{cc,1} \propto F^\infty_0$ and  
$X^{cc,2}\propto F^\infty_E-3 F^\infty_{{\bf P}^1}-X^{\infty,0}$. 
It follows that we can
turn on fluxes which lead to a superpotential $W= n_1 X^{cc,1}+n_2 X^{cc,2}$
and a scalar potential which is in leading order 
\beq
v=- a_1 {n_1^2 \over \log(|x_1|) }- a_2{n_2^2 \over \log(|x_2|) } ,
\eeq
where $a_1,a_2>0$.   
Note that we have chosen the flux vector such that it fixes the minimum of 
the potential in the  moduli space at complex codimension two.

\subsection{The Seiberg-Witten limit}

In the type II geometry, the  Seiberg-Witten $SU(2)$ theory 
emerges at the blow up in the moduli space, 
which resolves the tangency of the weak coupling divisor 
$y\propto e^{-S}=0$ and the generic conifold locus. 
The simplest situation where this geometry arises is for 
the  $2$ moduli $K_3$ fibration examples studied in \cite{CDGMK,HKTY,KKLMV}. 
A schematic picture of the moduli space for this type of models 
can be found in  Fig. 4 of \cite{CDGMK}.  As shown in Fig. 1 below 
we will use a slightly different 
resolution of the tangency of $\Delta_{con}$ and $y=0$ 
than refs. \cite{CDGMK,KKLMV}.
The advantage is that this resolution splits the monopole and the dyon 
point, which is important when writing down the superpotential. Furthermore 
the SW-monodromy group is embedded in the simplest possible way 
in the Calabi-Yau monodromy group. In \cite{Mayr} model independent general 
expressions for the scalar potential have been obtained at the 
Seiberg-Witten point. Here we will derive the scalar potential for a
specific model.

In particular, for the well studied degree 12 hypersurface 
in ${\bf P}(1,1,2,2,6)$ we have $C_{TTT}=4$, $C_{STT}=2$, zero 
otherwise, and $\int_W c_2 J_T= 52$, $\int_W c_2 J_S=24$ and $\chi=-252$. 
Here we denote the K\"ahler class measuring  the complex volume of 
the ${\bf P}^1$ basis of the $K_3$ fibration by $S$ and the square root of the complex 
volume of the $K_3$ by $T$. The integral basis is fixed  by (\ref{basis}).
The connection to the canonical large complex structure variables is given by 
the leading order relations $z_t\propto e^{-T}, z_s\propto e^{-S}$. We 
use the same rescaled complex structure variables as in 
\cite{HKTY}, $x=1728 z_t$ and $y=4 z_s$. 
The order two tangency between the conifold divisor 
$\Delta_{con}=\{\Delta^{con}_+\Delta^{con}_-=
\left((1-x)+x \sqrt{y} \right)\left((1-x)-x\sqrt{y}\right)= (1-x)^2-y x^2=0\}$
and the weak coupling singularity $W=\{y=0\}$  is
now resolved by blowing up which introduces an exceptional divisor 
$E\sim {\bf P}^1$, see the right-hand side of Fig. 1. Near $W\cap E$ 
the dimensionless  variables 
$w_1={x \sqrt{y} \over (1-x)}= {1\over {\tilde u}}$ and 
$w_2=(1-x)=\alpha' u +{\cal O}(\alpha'^2)$ 
are good coordinates. Here $\tilde u={u\over \Lambda^2}$ is 
dimensionless of order one 
and the identification with the Seiberg-Witten variables to 
first order in $\alpha'$ is dictated by the double 
scaling limit \cite{KKLMV} $\epsilon \rightarrow 0$ in 
\beq
y=e^{-S}=:(\alpha')^2 \Lambda^4 e^{-\hat S}=:\epsilon^4,\qquad
(1-x)=\alpha'u+{\cal O}(\alpha'^2)=\epsilon^2\tilde u+{\cal O}(\epsilon^4)\ . 
\eeq
The last equation implies in particular  
$\sqrt{\alpha'} \Lambda = {\Lambda\over M_{Str}}=\epsilon$. 
Since  $\tilde u e^{-S/2} \sim (1-x)$ and $(1-x)$, 
is proportional at weak coupling to the mass square of $W^\pm$  
in string units, the exponential relation in the double scaling limit  is a 
reflection of the renormalization group 
equation ${8 \pi^2 \over b_1 g^2(M_{Str}}=-\log({m_{W^\pm}\over M_{Str}})$ 
due to the exact one-loop 
$\beta$-function of ${\cal N}=2$ with coefficient $b_1=4$ for pure 
$SU(2)$ Yang-Mills theory.

\parbox{16cm} 
{

   \begin{center}
   \mbox{ 
             \epsfig{file=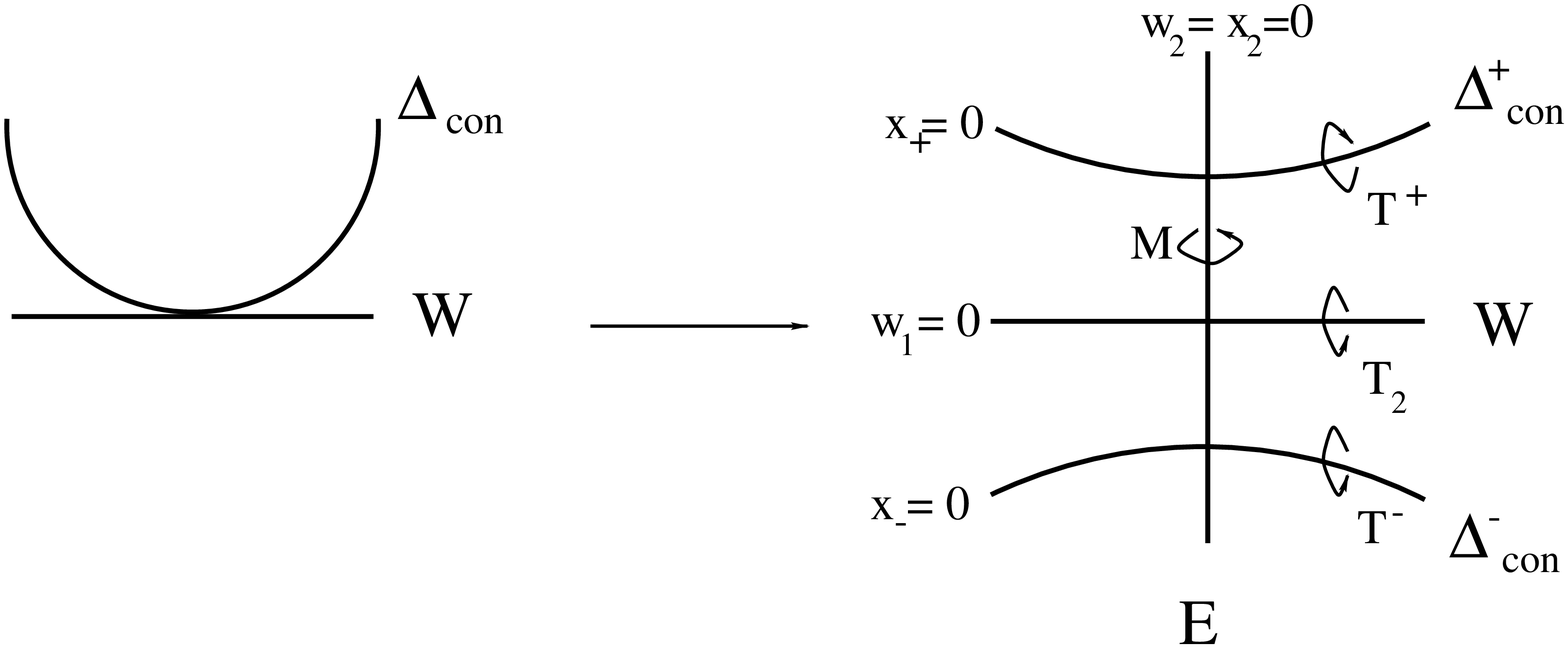,width=8cm}
}
   \end{center}
}
\centerline{{\bf Fig. 1} Blow up of the Seiberg-Witten point.} 

The Picard-Fuchs equation in \cite{HKTY} can be 
easily solved in the $(w_1,w_2)$ coordinates 
near $W \cap E$. This corresponds to the classical field theory limit
in sect. 3.4.1 (considered there for the $STU$ model. 
\begin{eqnarray}
\Xi^1_{u=\infty}=1+ {\cal O}((\alpha' u)^2)  
&=& 1-{5 \over 216}\left(1 + {1\over2}w_1^2\right)w_2^2+{\cal O}(w_2^3) \nonumber\\
\Xi^2_{u=\infty}=\alpha' u  + {\cal O}( (\alpha' u)^2 )   
&=& w_2-{77\over 108}\left(1+{1\over2}w_1^2\right)w_2^2 
+ {\cal O}(w_2^3)\nonumber\\
\Xi^3_{u=\infty}=\sqrt{\alpha'} a(\tilde u)(1 + {\cal O}(\alpha' u ))
&=&
{i\over \pi} \sqrt{w_2}(1-{1\over 16}w_1^2-{15\over 1024} w_1^4+\ldots )
+{\cal O}(w_2^{3/2}) \nonumber\\
\Xi^4_{u=\infty}= s(1+ {\cal O}((\alpha' u)^2)
&=&
{\Xi^1_{u=\infty}\over \pi i}\log(w_1 w_2)-{32\over 81}w_2^2+{\cal O}(w^3)\nonumber\\
\Xi^5_{u=\infty}= \alpha' u s(1+ {\cal O}(\alpha' u)) 
&=&
{\Xi^2_{u=\infty}\over\pi i}\log(w_1w_2)-{139\over 81} w_2^2+{\cal O}(w^3) \nonumber\\
\Xi^6_{u=\infty}=\sqrt{\alpha'} a_D(\tilde u)(1 +{\cal O}(\alpha' u) )  
&=&{\Xi_{u=\infty}^3\over\pi i}\log(w_1)+ {2 \sqrt{w_2}\over\pi^2} 
(1-{1\over 8}w_1^2+{\cal O}(w_1^4))+\kappa\Xi^3_{u=\infty},\nonumber\\ 
\label{basisatswinf}
\end{eqnarray}
with $\kappa={i\over \pi} (3 \log(2)-1)$. Classical gauge group enhancement 
to $SU(2)$ occurs at $a=0$ where the period $\Xi^3_{u=\infty}$ vanishes.
   
Near the conifold branch $\Delta^{con}_{+}\cap E$, 
$x_+={1-x\over x \sqrt{y}}-1= {\tilde u}-1$ and  $x_2=(1-x)=
\alpha' u+{\cal O}(\alpha'^2)$
are good coordinates.  Near the branch $\Delta^{con}_{-}\cap E$ we 
may chose $ x_-={1-x\over x \sqrt{y}}+1= {\tilde u}+1$ and  $x_2=(1-x)=
\alpha' u+{\cal O}(\alpha'^2)$. The leading terms of the above basis in the 
$(x_+,x_2)$ coordinates near $\Delta^{con}_+\cap E$ read\footnote{Note that 
$w_1^{1/2}w_2^{1/2}=(x_++1)^{1/2}w_2^{1/2}=\sqrt{\alpha'}\Lambda $, 
has to be factored out from the solutions on the
right to obtain the Seiberg-Witten 
expansions.}
\begin{eqnarray}
\Xi^1_{mon}=1 + {\cal O}( (\alpha' u)^2 )  
&=& 1-{5\over144}\left(1+{2\over3}x_+\right)x_2^2+ {\cal O}(x_2^3) 
\nonumber\\
\Xi^2_{mon}=\alpha' u +{\cal O}((\alpha' u)^2)   
&=& x_2-{77\over72}\left(1+{2\over 3}x_+\right)x_2^2 
+{\cal O}(x_2^3)
\nonumber\\
\Xi^3_{mon}=\sqrt{\alpha'} a(\tilde u)(1 + {\cal O}(\alpha' u ))
&=&
{i \Xi_{mon}^6 \over 2 \pi }\log(x_+)
+{\sqrt{2x_2}\over i\pi^2}(1-{46\over81} x_2
+ {\cal O}(x^2))+\delta \Xi_{mon}^6
\nonumber\\
\Xi^4_{mon}= s(1+ {\cal O}((\alpha' u)^2)
&=&
{\Xi^1_{mon}\over\pi i}\log(x_2)-{1\over\pi i}\log(1-x_+)+{\cal O}(x_2^2)  
\nonumber\\
\Xi^5_{mon}= \alpha' u s(1+{\cal O}(\alpha' u)) 
&=&
{\Xi^2_{mon}\over\pi i}\log(x_2)-{x_2\over\pi i}\log(1-x_+)+{\cal O}(x_2^2) 
\nonumber\\
\Xi^6_{mon}=\sqrt{\alpha'} a_D(\tilde u)(1 +{\cal O}(\alpha' u) )  
&=& {1\over \sqrt{2} \pi} \sqrt{x_2}(x_++{9\over 32} x_+^2+ 
{75\over 256} x_+^3+ \ldots )+{\cal O}(x_2^{3\over 2}), 
\nonumber\\ 
\label{basisatconifold}
\end{eqnarray}
with $\delta=1 +{3 i (1-2 \log(2))\over \pi}$. Note that $s= 2 \pi i S\propto 
{2 \pi i \over g^2}$ and $-S=\log(y)$.

In this simple model we  can give, at least numerically, a complete account 
of how the periods in the Seiberg-Witten field theory limit are related to 
the ones in the large complex structure basis, in which our flux 
choices are made, by calculating the transformation matrix\footnote{Explicit results for this 
Calabi-Yau manifold had already been obtained by W. Lerche and P. Mayr (unpublished notes).} 
$\Xi_{u=\infty}=\Xi_{mon}=N \Xi_\infty$, where the basis at infinity is 
as in (\ref{basis}), 
$\Xi_\infty\propto(1,t,s,\partial_s F,\partial_t F,
2 F-s \partial_s F-t\partial_t F)$
\begin{eqnarray}
N=\left(\matrix{
0   & i A_+ B  & 0   &   {A_- B\over 2} &     0    &      0  \cr 
0   &   i  B   & 0   &    {B\over 2}    &     0    &      0  \cr  
1   &   0      & 0   &    -{1\over 2}   &     0    &      0   \cr
0   &v_2 +iA_-B&A_-B &{A_-B\over 2}+iv_1&-{i A_+B\over 2}&-{A_-B\over 2}   \cr
0   &v_3+ i  B &  B  &{B \over 2}+ i v_4&{iB\over2}&{B\over 2} \cr
0   & 0        &  0  &      0           & 0        &     1       }
\right)\label{analyticconta}
\end{eqnarray}
with  $A_\pm={1\over 36 \pi^4} (5 \pi^4\pm 12 \Gamma^8\left(3\over 4\right))$, 
$B=-{\pi^3 \sqrt{3}\over 
\Gamma^4\left(3\over 4\right)}$, $v_1\approx -4.0767326$, 
$v_2\approx -16.409393$, 
$v_3\approx -69.6002844$ and 
$v_4\approx -8.61884321$. From the third and last line in 
$N$ one sees that the periods, 
which contain the Seiberg-Witten 
periods in the normalization\footnote{Note that $\Lambda$ is 
rescaled by a numerical 
factor of order one, $\Lambda={\pi \over i \sqrt{2}}\Lambda_{sw}$.} 
$(a\sim {1\over 2}\Lambda_{sw}\sqrt{2\tilde u},
a_D\sim {2 i\over\pi}a\log(\tilde u))$, 
have intersection $1$ so we can make them dual in a symplectic 
basis, but because of the entry $-{1\over 2}$ not in an 
integral symplectic basis.

The monodromy in the above basis around $x_+=0$ can be 
identified directly with the Seiberg-Witten monopole
monodromy $\tilde M_{(0,1)}$ while the combination of 
monodromies around $w_2=0$ and $w_1=0$ give  
the the Seiberg-Witten monodromy around infinity  
$\tilde M_\infty=M T_2^{-1}$
\beq
\tilde M_\infty=
\left(\matrix{
1&0&\phantom{-} 0&0&0&0\cr
0&1&\phantom{-}  0&0&0&0\cr
0&0&-1&0&0&0\cr
0&0&\phantom{-} 0&1&0&0\cr
0&0&\phantom{-} 0&0&1&0\cr
0&0& \phantom{-}4&0&0&-1}\right)\qquad 
\tilde M_{(0,1)}=\left(\matrix{
1&0&0&0&0&\phantom{-}0\cr
0&1&0&0&0&\phantom{-}0\cr
0&0&1&0&0&-1\cr
0&0&0&1&0&\phantom{-}0\cr
0&0&0&0&1&\phantom{-}0\cr
0&0&0&0&0&\phantom{-}1}\right)\ .
\eeq
It is a check on $N$ that these are integral in the basis (\ref{basis}) 
and can be identified with $\tilde M_{(0,1)}=T$ and 
$\tilde M_\infty=A^{-1}T A T$ in
the notation of \cite{KKLMV}. Similarly, the monodromy around $ x_-$ gives  
$\tilde M_{(-1,1)}=\tilde M^{-1}_{(0,1)}\tilde M_\infty$.

Let us first discuss the superpotential which arises in the classical field theory limit, 
\beq
W=n \, \Xi_{u=\infty} \sim  { n \sqrt{\alpha'} a } \ ,
\eeq
which vanishes at the point of classical gauge group enhancement.
Using the third row of the matrix $N$, we see that 
\beq
\Xi^3_{u=\infty}=X^0_\infty- {i \over 2} {\partial F\over \partial S}\   
\eeq
and hence the superpotential becomes 
\beq
W= n (1- {i\over 2} {\partial F\over \partial S } )= {n\over 2} ( 1-T^2)\ .
\eeq 
This superpotential matches exactly with  the superpotential in 
(\ref{supals}) after setting $e_0=- 2 m^1=n$ and $e_2=0$.

Let us now go to the point where the monopole becomes massless.
We first discuss the field theory expectations and  
assume as in \cite{TayVa} that there is flux 
such that the field theory superpotential behaves in 
leading order at the Seiberg-Witten point 
as $W\sim m u$ \cite{sw}. This corresponds to a mass term 
for the adjoint scalar, which breaks ${\cal N}=2$ to ${\cal N}=1$. 
Roughly speaking, such a potential should be generated  by a  
flux that has $m$ ``units'' on $\Xi^2_{mon}$, i.e. $W=m X^2_{mon}$. 
The superpotential has dimension 
three and since the parameters $x_+,x_2$ are dimensionless it 
reads in natural units as 
\beq 
W={m\over (\alpha')^{3\over 2}} X^2_{mon}
\sim {m\over (\alpha')^{3\over 2}} x_2 + {\cal O}(x^2)\ . 
\eeq 
Under the double scaling limit it behaves hence as 
$W\sim m M_{str} \Lambda^2 \tilde u$.   

In the ${\cal N}=1$ 
field theory one expects $h$ vacua with a mass gap, 
where $h$ is the dual Coxeter number of the gauge group.   
At each vacuum there is a superpotential 
\beq
W_k=w^k e^{-S/h}\ , 
\eeq
where $w^h=1$, $k=0,\ldots,h-1$ and $S={1\over g^2}$. 
Indeed we find that $\Xi_{mon}^2\sim x_2$ 
and from the period $\Xi_{mon}^4$ we learn that $x_2\sim e^{-{S\over 2}}$, so that the string 
embedding delivers precisely the right behavior of the superpotential.   

The main  issue will be the  degeneration of the 
factor $e^K G^{x_i\bar x_j}$ and whether the 
the scalar potential drives the theory towards the Seiberg-Witten point. With the inverse of $N$ 
\begin{eqnarray}
N^{-1}=
\left(\matrix{
  C             &    -C_+        &  {1 \over 2}      &      0        &      0                 &    0      \cr
   iC          &-i C_-          &      0            &      0        &       0                &       0 \cr
   C+i u_3      &     C_++i u_4  &         0         &      C      &          -C_+         &  {1\over 2}\cr
    2 C         &     -2 C_+     &           0       &       0       &      0                &      0 \cr 
  u_1+2 i C     &      u_2-2iC_- &       0           &   -2i C     &   2 i C_-             &      0\cr  
      0           &      0         &       0           &      0        &        0              &      1}
\right)\label{analyticcont}
\end{eqnarray} 
where $C={\sqrt{3} \pi \over 2 \Gamma({3\over 4})^4}$ and 
$C_\pm={\sqrt{3}  \over 72 \Gamma({3\over 4})^4}(5 \pi^4\pm 12
\Gamma({3\over 4})^8)$, $u_1\approx 5.5157560$, 
$u_2\approx -.05616975$, $u_3\approx  .1051578$  and $u_4\approx  .263801$, 
we
find that the K\"ahler factor $e^K$ 
diverges at $E\cap \Delta^{con}_+$
as
\beq
e^K={1\over 2} {\pi \over \pi (2 u_3+ u_1)-4 C\log(|x_2|)}- 
{{\rm Re} (x_2) a- 4 C^2 \pi {\rm Re} (x_+)\over 
(\pi (2 u_3+ u_1)-4 C\log(|x_2|))^2}+{\cal O}(x^{2}) \ , 
\eeq
with $a=\pi (2 u_3 C_-- C(2u_4+u_2)+C_+ u_1)$.
To leading order the inverse metric is 
\beq
G^{x_i \bar x_j}=  
4 \pi^2 C  {\pi (2 u_3+ C)  - 4 C \log( |x_2|) \over 6 \log(2)-\log( |x_+|)  }\left(\matrix{ 
{1 \over |x_2|} &-\sqrt{x_2\over \bar x_2}\cr
-\sqrt{\bar x_2\over x_2}& |x_2| }\right) \ .
\eeq   

The scalar potential, to  
leading order in $(x_+,x_2)$, is
\beqa
v=& m^2 {2 \pi^3\over (\alpha')^2}
\Biggl(|x_2|^2\Bigl[\pi^2(u_1+2u_3)^2+ 16 C^2 \log(|x_2|)^2 -8\pi C(2 u_3 +1)\log(|x_2|)\Bigr]- \cr 
  & 4 ({\rm Im}(x_2))^2\Bigl[2 C(2C+\pi (u_1+2 u_3))-8 C^2\log(|x_2|)\Bigr]\Biggr)\times\cr 
  &  \left(|x_2|\Bigl(6 \log(2) - \log(|x_+|)\Bigr) \Bigl(\pi (u_1+2 u_3)-4C  \log(|x_2|)\Bigr)^2 \right)^{-1} \ . 
\eeqa

We have plotted $v$ in Fig. 2.
 
\parbox{16cm} 
{

   \begin{center}
   \mbox{ 
             \epsfig{file=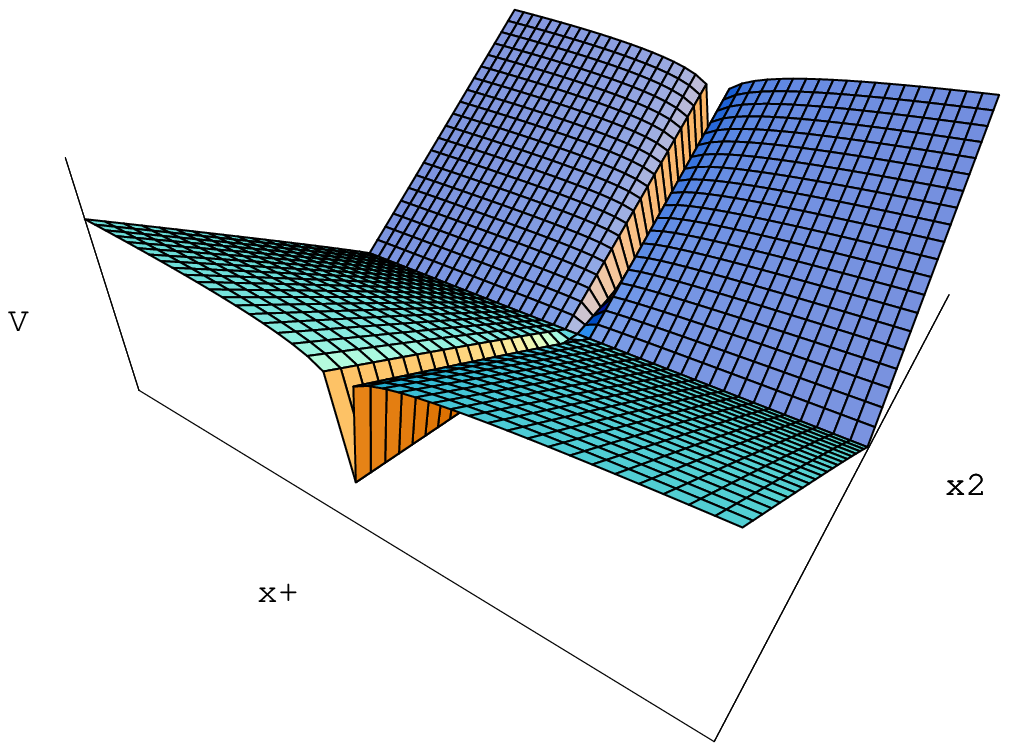,width=8cm}
}
   \end{center}
}
\centerline{{\bf Fig. 2} Potential near the Seiberg-Witten point $x_+=x_2=0$ for the
flux along $\Xi^2_{mon}$.} 

The potential exhibits to this 
order two flat directions at $x_+=0$ and at $x_2=0$ 
along which $v=0$. Similar to the conifold case we 
find that there is a supersymmetric vacuum despite
the fact that ${\dd W \over \dd x_2}\neq 0$ at $x_2=0$ 
due to the degeneration of the metric and the $e^K$ factor.
Due to the $1/\log(|x_+|)$ term the first derivative in 
the $x_+$ direction becomes infinite at $x_+=0$. 
So the potential drives the theory strongly to $x_+=0$. 
At $x_+=0$ the flatness in the $x_2$ is lifted
at order $|x_2|^2$. We find at this order in $|x_2|^2$ a term\footnote{There are further 
terms subleading in $\log(|x_2|),\log(|x_+|)$. E.g. the $e^K W \bar W$ term 
contributes at this order with ${1\over 2} \left(\pi \over {\pi (2 u_3+u_1)-4 C \log(|x_2|)}\right) |x_2|^2$.}  
\begin{equation}
\lim_{x_+=0,x_2\rightarrow 0} v=-m^2
{\pi |x_2|^2\over 2 (\alpha')^2 C^2} \log(|x_2|)= -m^2
\Lambda^4  {\pi |\tilde u|^2\over 2 C^2} 
(\log(|\tilde u|)+{\rm const.} ) \, .
\end{equation} 
It is worth noting that in the double scaling limit
this first nonzero contribution of the expansion of the potential has 
the expected scale $\Lambda^4$ of the field theory potential.

Now let us discuss in more detail the transition from the period 
$\Xi_{mon}^2$ to the periods $\Xi_\infty$ at infinity, which is
necessary since the integral fluxes are defined only with respect
to $\Xi_\infty$.
The fact, which allowed for the precise identification of the above  
flux,  is that there is a unique period  $\Xi_{mon}^{2}$ that behaves like 
$x_2+{\cal O}(x_2^2)=\alpha' u+{\cal O}((\alpha' u)^2)$ 
in the double scaling  limit.
The analytic continuation (\ref{analyticcont})
of this period to infinity (\ref{basis}) reads
\begin{equation}
\Xi_{mon}^{2}= \, B  \, T + {i\over 2}\,  B\, {\partial F\over \partial S} .
\label{swcont}
\end{equation}
The flux which leads to  $W\sim m  u$ term hence corresponds to a  electric
charge
$e_2$ and a magnetic charge $m^1$. 
The irrational number $B$ can be absorbed in the definition of $m$, 
namely the integer fluxes are $m^1={1\over 2}mB$ and $e_2=-imB$. 
Note that the relative factor between the flux vector entries is $i/2$. 
This means that the superpotential $W=m u$ 
cannot be generated by a Ramond flux alone. 
Specifically, whereas $m^1$ is a real Ramond flux, 
the electric flux $e_2$, which corresponds to the field $T$,
is purely imaginary and hence is generated by a NS flux, where we have
chosen the complex field $\tau$ to be imaginary, $\tau=i$. 
    
This choice of fluxes can be compared to previous discussions in the
literature on this issue. First,
the identification of the flux direction as 
${\partial F\over \partial S}\sim u$ in \cite{TayVa}
ignores the mixing by the analytic continuation. Second, 
the above identification has no 
zero-brane charge as in \cite{Mayr}. This is explained by the fact 
that the basis used here differs by a integral symplectic transformation 
relative to the one used in \cite{Mayr}, i.e. our charge definitions are 
different.    

If we just turn on the flux $m^0$ the leading behaviour of 
the scalar potential is 
\beq 
v=- { \pi \over  2 (\alpha')^2 (   \log(|x_+|) + 6 \log(2))}\ .
\eeq
This drives the theory to the conifold line, where the potential 
vanishes .  

We have also computed the leading $x_2$ correction; its expression is
very complicated. It is interesting that
in higher order the $x_2$ direction is lifted so that the theory 
is driven towards the Seiberg-Witten point. We have plotted the potential
to ${\cal O}(x^2)$ in fig. 3.  
 
\parbox{16cm} 
{

   \begin{center}
   \mbox{ 
             \epsfig{file=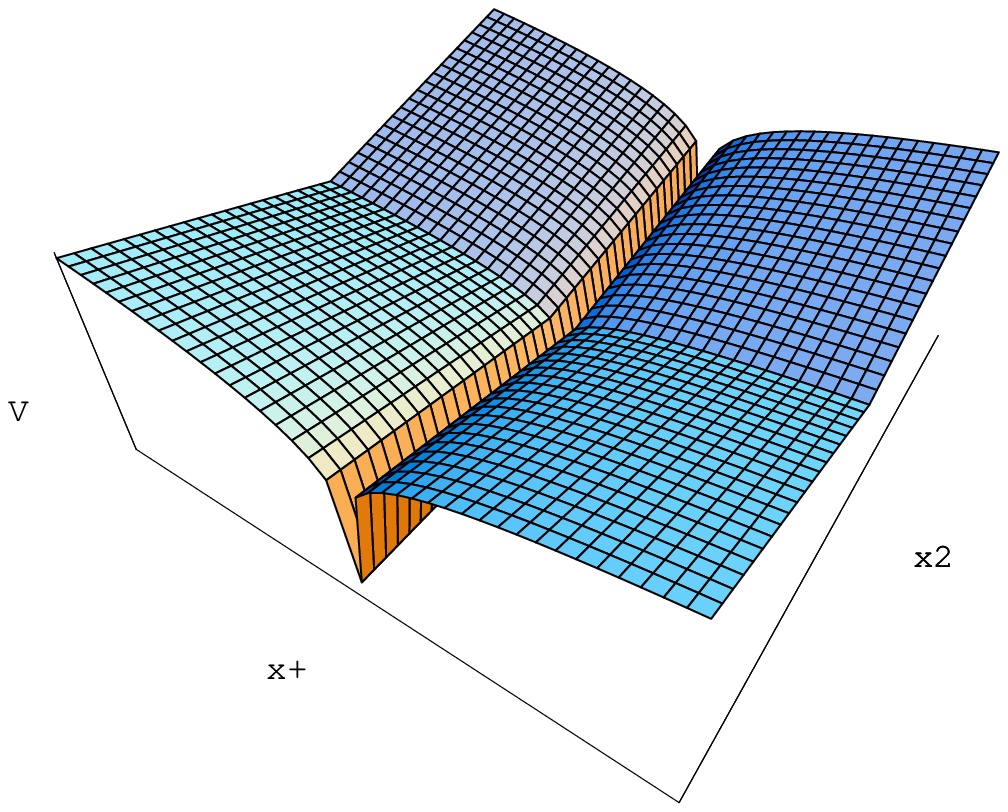,width=8cm}
}
   \end{center}
}
\centerline{{\bf Fig. 3} Potential to order ${\cal O}(x^2)$ 
near the Seiberg-Witten point 
$x_+=x_2=0$ with flux on $F_0$.} 

As already explained, besides going to the $u$-plane, the Seiberg-Witten
limit also requires going to the conifold limit, 
$F_0\rightarrow 0$, where a non-perturbative monopole hypermultiplet
becomes massless. Therefore we can also turn on the corresponding 
flux $m^0$. (We could also turn on fluxes along 
$X^3_{mon}$, $X^5_{mon}$, which also vanish.) Performing these two limits 
and including the mass term for the
light monopole hypermultiplet $\phi_M$ and $\tilde\phi_M$, 
the  superpotential becomes
\beqa W=
-m^0 F_0+  m\biggl( {i}{\partial F\over\partial S}+
2 T\biggr)
+F_0\phi_M\tilde\phi_M \, .\label{supmon}
\eeqa
Minimization of the corresponding potential will relate the $vev$
of the monopole hypermultiplet $\phi_M$ to the vector moduli fields
$u$ and $S$.

Let us briefly comment on the possibility of partial supersymmetry
breaking from ${\cal N}=2$ to ${\cal N}=1$ at the Seiberg-Witten
point. As explained 
in \cite{TayVa}
partial supersymmetry breaking is possible in the
rigid field theory limit after freezing out the $S$-field. From the
local supergravity perspective partial supersymmetry breaking seems
possible since the flux vector $e_2$ is imaginary, as soon as we freeze
out the hypermultiplet $\tau$. Then the difference in the mass eigenvalue
of the two gravitini is proportional to $2mBT$. But if we treat
$\tau$ as a dynamical field, the minimization of the potential with
respect to $\tau$ will make partial supersymmetry breaking impossible, as
explained before.

\subsection{The strong coupling limit and the r\^ole of the $\beta$-function.}

The heterotic dilaton $s$ corresponds in the dual $K3$ fibered Calabi-Yau spaces
to the complexified volume of the ${\bf P}^1$ base. 
Its dependence on its mirror dual 
complex structure modulus is governed by an universal differential equation 
($\theta=y {\dd \over \dd y}$, $y=4 z_s$) (see e.g. \cite{HKTY}
\beq
\left[\theta^2- {y\over 4} \theta (\theta+{1\over 2})\right]\Xi=0 \  ,
\label{degl1} \eeq
with three regular singular points at $y=(0,1,\infty)$. Eq. 
(\ref{degl1}) implies, that if the other moduli of the Calabi-Yau $z_i$ 
are set to $0$, the complexified volume $s$ is given by 
\beq 
s= {1\over \pi }  \arctan (\sqrt{y -1}) \ .
\eeq
In particular at $y=1$ the period ratio $s$ vanishes  
with $w_{s}=1-y$  as $s={1\over \pi} \sqrt{w_s} \sum_{n=0}^\infty {w_s^n\over 2 n +1}$.  
In the heterotic string this corresponds to the strong coupling limit.
We therefore refer to the $w_s=1-y=0 $ locus in the moduli 
space as strong coupling divisor. Inside the Calabi-Yau one finds a divisor 
$F$ which is a rational fibration with the ${\bf P}^1$ as fiber and a 
genus $g$ curve  $C_g$ as base. 
If the fiber ${\bf P}^1$ vanishes, $F$ collapses to $C_g$ and  branes wrapped 
on ${\bf P}^1$ yield in type IIA compactifications charged massless 
vector multiplets, which complete the $U(1)$ vector multiplet 
$s$ to a $SU(2)$ vector multiplet. 
The square root branch cut $\sqrt{w_{s}}$ will generate the Weyl 
reflection as monodromy. Holomorphic one-forms of $C_g$ lead to additional 
matter multiplets, likewise in the adjoint representation \cite{strong}.
We will investigate superpotentials at the strong 
coupling singularity. An example is the $X_{12}(1,1,2,2,6)$ 
model discussed in the last section. The curve $C_g$ is the 
vanishing locus of the  first two variables and is hence represented 
by $X_6(1,1,3)$. By the adjunction formula 
$g=(2-\chi)/2=1-{(1+J)^2(1+3 J)6J\over 2\cdot 3 (1+6J)}|_{J^2}=2$. With 
two hypermultiplets in the adjoint representation the coupling 
grows with the scale.
 
At the point $(x=1728 z_t ,y)=(0,1)$
we have in the $(z_t,w_s=1-y)$ coordinates the solutions 
\begin{eqnarray}
\Xi^1_{s}&=& 1+{5\over 72} z_t+{\cal O}(2) 
\nonumber\\
\Xi^2_{s}&=&{1\over 2\pi i}\left(\Xi^1_s\log(z_t)+{1\over 2}z_t
+{1\over 2}\log(1-w_s) +{\cal O}(2)\right)
=:{1\over 2\pi i}\left(\Xi^1_s\log(z_t)+\Sigma_1\right)
\nonumber\\
\Xi^3_{s}&=&
{i \over  \pi }\sqrt{w_s}\left({\arctan(\sqrt{-w_s})\over i \sqrt{w_s}} 
+{5\over 216} z_t w_s+ {\cal O}(2)\right)=a(w_s)+{\cal O}(z_t) 
\nonumber\\
\Xi^4_{s}&=&{\Xi^3_{s}\over 2 \pi i} \log(z_t w_s)
-{\sqrt{w_s}\over 2 \pi^2}({5\over 18} w_s +{\cal O}(2))=a_D+{\cal O}(z_t)  
\nonumber\\
\Xi^5_{s}&=& 
{1 \over 2\pi^2}\left(\Xi_s^1\log(z_t)^2
+2\Sigma_1\log(z_t)-w_s+{13\over 18} x + {\cal O}(2) \right)=: 
{1 \over  2 \pi^2}\left(\Xi_s^1 \log(z_t)^2 +2\Sigma_1\log(z_t) 
+\Sigma_2 \right)
\nonumber\\
\Xi^6_{s}&=&{2 \over 3 ( 2 \pi i)^3}( \Xi^1_{s}\log(z_t)^3+
3 \Sigma_1\log(z_t)^2 + 3\Sigma_2\log(z_t) -6 w_s + {\cal O}(2)). 
\nonumber\\ 
\label{basisatstrongcoupling}
\end{eqnarray}

This basis is related to the basis at $z_t=z_s=0$ by the matrix 
$\Pi_\infty=N_{str}\Pi_s$ with (cf. footnote 16)
\begin{eqnarray}
N_{str}=
\left(\matrix{
1     &   0   &   0  &   0   &  0   &  0   \cr 
i b_0 &   1   &  -{1\over 2} &  0 & 0&  0  \cr  
0     &   0   &   1  &   0   &  0   &  0   \cr 
2 b_1 &-4 i b_0 & 0  &   1   &  0   &  0   \cr
b_1   &  -2i b_0& ib_3 &{1\over 2}  &  1&0 \cr
-i b_2& b_4   &  0   & -ib_0 &0     & 1    \cr
}
\right)\ , \label{infinitytostrongcouping}
\end{eqnarray}
Here $b_0={\log(2)\over 2 \pi}$ and 
$b_1\approx 1.09550$,
$b_2\approx 1.00245$,
$b_3\approx .20799$ and
$b_4\approx 2.14233$.

We consider a flux on the vanishing period $\Xi_s^3=s$, 
which generates a superpotential
\beq
W= n s\ 
\eeq  
and find the potential in leading order in $(z_t,w_s)$  
\beq
v={n^2 \pi \over (\alpha')^2 ( 2 \pi b_3-2\log(|w_s||z_t|) -2)}+{\cal O}(1)\ .
\eeq 
This potential exhibits a similar logarithmic behaviour as the one at the 
conifold or the pure $SU(2)$ point. 
Its value is zero at $w_s=0$ and at $x=0$. 
This value represents a local minimum and the theory is strongly 
(by an infinite slope) attracted to $\lbrace w_s=0=x=0\rbrace$. 
The $\log(|w_s| |z_t|)$ dependence is due to the 
non-vanishing of the $\beta$-function
of the gauge coupling constant. We thus expect a qualitative 
new behaviour of the scalar potential  
at conformal points where the $\beta$-function vanishes.

\subsection{The conformal points}

We can study the situation with vanishing $\beta$-function 
in the $X_{24}(1,1,2,8,12)$ $K3$ fibration. 
Here $C_g$ is obtained, as before, by  
setting to  zero the first two variables, which yields $X_{12}(1,4,6)$. 
Note that this curve has the $Z_2$ singular point $X_6(2,3)$.
The Euler number is obtained by combining
the adjunction formula with the Riemann-Hurwitz formula: 
$\chi(C_g)={(1+J)(1+4 J)(1+6 J)12J\over 4\cdot 6(1+12J)}|_{J^2}
-1/2+1\cdot 1=0$. Hence there is $g=1$
hypermultiplet and the spectrum is conformal.        

Recall that the above CY space $M$ with Euler number $\chi=-480$ 
exhibits several fibration structures. On the one hand it is 
an elliptic fibration 
over the Hirzebruch surface ${\bf F}_2$, which is itself a rational 
fibration with fiber ${\bf P}^1_U$ and base ${\bf P}^1_S$. 
At the same time the CY space is a $K3$ fibration over the same 
base, hence the identification of the heterotic dilaton with $S$.  
We have as K\"ahler classes $S$, 
the size of the base ${\bf P}_S^1$, $U=\tilde U-\tilde T$ the size of the 
fiber ${\bf P}_U^1$ and $T=\tilde T$ the size of a curve in the elliptic fiber.
Hence we have the nonvanishing 
intersections\footnote{Quantities with the tilde refer here to 
the heterotic spacetime moduli of $T^2$ in the compactification on
$K3\times T^2$, i.e. $\tilde U$ is the complex structure modulus 
of $T^2$ and $\tilde T$ its K\"ahler modulus.} 
$C_{TTT}=8$,       
$C_{STT}=2$,       
$C_{UTT}=4$,       
$C_{STU}=1$,      
$C_{TUU}=2$, $\int c_2 J_T=92$, $\int c_2 J_S=24$ and $\int c_2 J_U=36$. 
This fixes the integral basis. The mirror manifold $W$ has the following 
discriminant (we have rescaled 
$x=432 z_t$,
$y=4 z_s$,
$y=4 z_u$)  
\beq
\Delta_{s}\Delta_{A} \Delta_{B}=(1- y)[(1-z)^2 - y z^2]
[((1-x)^2-z)^2-y z^2] 
\eeq 
which consists of three factors. 
$\Delta_A,\Delta_{B}$ are conifold loci. 
At $z=1$, $y=0$,  $\Delta_{B}\neq 0$ there is a weakly 
coupled $SU(2)$ without matter \cite{KKLMV}. 
We will be interested in the  $SU(2)$ with an adjoint hypermultiplet, 
i.e. with a conformal spectrum, which arises at $y=1$,  
$\Delta_{A}\neq 0\neq \Delta_B$, e.g. at $y=1$, $x=z=0$.   
To understand the r\^ole of the hypermultiplets it is interesting 
to contrast this model with the more generic realization of the 
$X_{24}(1,1,2,8,12)$ CY, which is given by an elliptic 
fibration over ${\bf F}_0$. 
As discussed in \cite{MVII} both models are in the same moduli space. 
However in the $X_{24}(1,1,2,8,12)$ model the scalars in one 
hypermultiplet\footnote{This is the hypermultiplet related to 
the three-cycle which consists of the base ${\bf P}^1$ and a one-cycle
on the elliptic fiber.}
are set to a 
special value, which allows for the fibration structure of the divisor $F$.
The linear reparameterisation at large coupling 
$T\to T$, $S\to S$, $U\to U-S$ relates 
the classical couplings.     
For the mirror of the ${\bf F}_0$ fibration 
the discriminant only factorizes into two factors:
\beq
\Delta=\Delta'_{A} \Delta'_{B}=[(1-y)^2+(1-z)^2-1+y z]
[(1-x)^4-2 x^2(1-x)^2(y+z)+x^2(y^2+z^2)] 
\eeq 
where $x,y,z$ are the coordinates on moduli space. 
Here we have a weakly coupled $SU(2)$ without matter at 
$z=1$, $y=0$, $\Delta'_B\neq 0$ (e.g.  for $x=0$) 
and at $y=1$, $z=0$, $\Delta'_B\neq 0$ (e.g. for $x=0$)  
\cite{KKV}. We can hence understand the 
deformation in the hypermultiplet moduli space away 
from the $X_{24}(1,1,2,8,12)$ Calabi-Yau as giving 
mass to the hypermultiplet in the adjoint at $y=1$, $x=z=0$.

\parbox{16cm} 
{

   \begin{center}
   \mbox{ 
             \epsfig{file=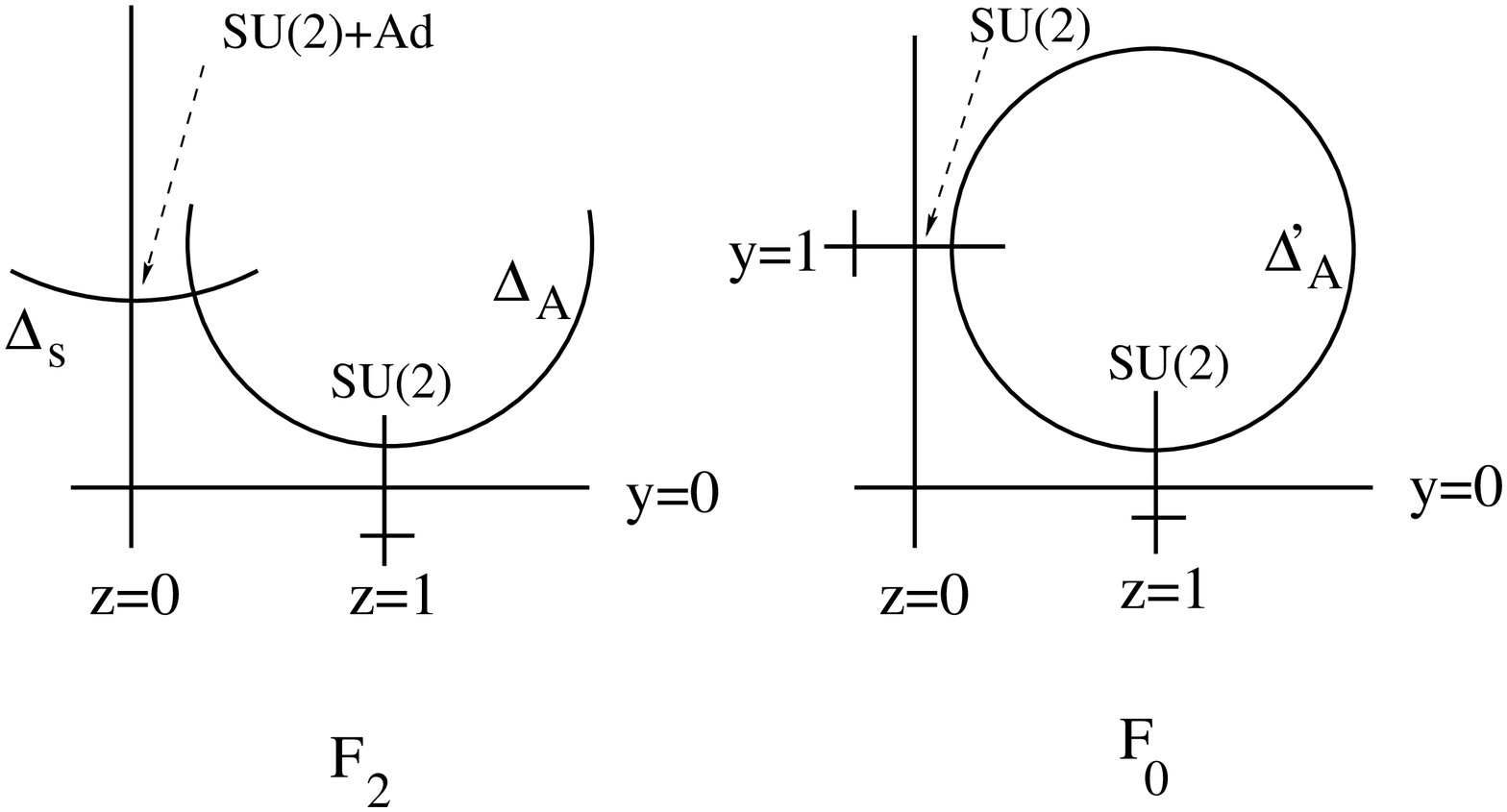,width=8cm}
}
   \end{center}
}
\centerline{{\bf Fig. 4} Slice of the quantum moduli space of the elliptic 
fibrations over ${\bf F}_2$, ${\bf F}_0$ at $x=0$ (local limit).} 

Let us now discuss the solution at $y=1$, $x=z=0$. As in equation 
(\ref{basisatstrongcoupling}) we have a unique solution 
$\Xi_s^1=1+{\cal O}(1)$. Furthermore the periods
$\Xi_s^2$,
$\Xi_s^4$,
$\Xi_s^5$,
$\Xi_s^6$ and
$\Xi_s^8$ have the same logarithmic behaviour in $\log(z_t)$ and 
$ \log(z_u)$ as (\ref{basis}). We can fix them uniquely by setting 
as many of the leading terms in the pure power series 
as possible to zero\footnote{We set the highest powers of 
$z_t,w_s,z_u$ in this order to zero.}. Let us give 
the  remaining solutions
\begin{eqnarray}
\Xi^3_{s}&=& {i\over\pi}\sqrt{w_s}({\arctan(\sqrt{-w_s})\over i\sqrt{w_s}}
+60 z_t +{\cal O}(2))=a+{\cal O}(z_t,z_u)) 
\nonumber\\
\Xi^7_{s}&=&{X_s^3\over 2 \pi i}\log(z_t)-{\sqrt{w_s}\over 2 \pi^2 }
(312 z_t+{\cal O}(2))=a_D+{\cal O}(z_t,z_u)) 
\label{basisatstrongcouplingII}
\end{eqnarray}
The crucial difference to (\ref{basisatstrongcoupling}) 
is that  the coordinate 
$w_s$ does not appear in the logarithm of the $a_D$ period.  
We can compute analytically the
whole transformation matrix except one numerical constant 
\begin{eqnarray}
N_{str}=
\left(\matrix{
1   &  0  &  0  &  0    &   0     &  0   &   0  & 0  \cr 
0   &  1  &  0  &  0    &   0     &  0   &   0  & 0  \cr 
0   &  0  &  1  &  0    &   0     &  0   &   0  & 0  \cr 
i A &  0  &  -{1\over 2}&   1     &  0   &   0  & 0 & 0  \cr 
B_- &- 4iA& 0  & -2 iA &    1     &  0   &   0  & 0  \cr 
1   &-iA  & 0  &   0    &   0     &{1\over 2}&1 & 0  \cr 
2   &-2 iA& 0  &   0    &   0     &  1   &   0  & 0  \cr 
C   &B_+  & 0  &   2    &   0     & -iA  &   0  & 1  \cr 
}
\right)\ , \label{infinitytostrongcoupingII}
\end{eqnarray}
where $A={\log(2)\over 2 \pi}$, $B_\pm={23\over 6} 
\pm {a\over (2 \pi i)^2}$, $a=.4804525$ and 
$C={63 i\over 2}{\zeta(3)\over \pi^3}+2iA$. We see from 
this matrix that the flux generating 
\beq 
W= n \Xi_s=n S
\eeq
is an integral Ramond-Flux w.r.t. the basis at large volume. 

The scalar potential is extremely complicated. The behaviour near 
$z_t=z_u=w_s=0$ is determined by   
\beq 
v \sim {p^0_8 + p^1_8 \sqrt{|w_s|} + p^1_9\over  
q_9 + q_{10}\ , }
\eeq
where $p^i_d$ and $q_d$ are generic homogeneous polynomials in 
$(\log(|z_t|),\log(|z_u|))$ of the indicated degree $d$.  
In particular the leading logarithmic term 
\beq 
v\sim - 2 \pi { 4 \log(|z_u|) + 9 \log(|z_t|)\over 
\log(|z_u|)^2 + 6 \log(|z_t|)  \log(|z_3|)+9\log(|z_t|)^2}\ 
\eeq  
is independent of $w_s$. The potential has a minimum at $z_t=0$ 
as well as at $z_u=0$ with vanishing energy. For fixed values 
of $z_t,z_u$ the potential can be analyzed to 10-th order in 
$\sqrt{w_s}$ and we found a series $v\sim \sum_{i=0} a_i(z_t,z_u) (|w_s|)^{i\over 2}$ 
whose coefficients $a_i$ go to zero if $z_t$ or $z_u$ goes to zero.
   
Similarly to the situation above (\ref{basisatstrongcouplingII}), where two 
periods vanish with a square root, it was found in \cite{Klemm:1996hh} at the point $w=0$, 
where tensionless strings arise, that two dual periods occur which start with 
$w^{1\over 6}$ and $ w^{5\over 6}$. Therefore we expect for fluxes aligned with a vanishing direction 
a similar behavior of the scalar potential in the direction of $w$, i.e. a  
power series in $w^{n\over 6}$ with $n\in {\bf Z}$.

\section{Type II Vacua with Non-vanishing Ramond and NS Fluxes}

\setcounter{equation}{0}

In sect. 3.7 on the discussion of the Seiberg-Witten limit NS-fluxes
have already emerged. We now want to study in little more detail
the possible effects of non-vanishing NS-fluxes for the vacuum
structure of type IIB strings. Since the superpotential is covariant
under the type IIB $SL(2,{\bf Z})$ duality symmetry (see eq.(\ref{tausutr})),
the scalar potential is invariant under this symmetry. Hence we expect
to find non-trivial minima, where the field $\tau$ is stabilized.
The following discussion will be performed in the $STU$ model,
where we can solve the minimization conditions explicitly. 
As we will see now, some of our previous conclusions concerning the
vector multiplets are modified by the presence of NS fluxes. 
In particular it seems
possible that the contribution of the R fluxes is precisely balanced
by the NS fluxes, such that supersymmetric minima of the scalar potential
are possible at non-degenerate points in the CY moduli space, i.e.
at finite value for the $S$-field.
It is worth pointing out that in the example we are going
to discuss the NS and Ramond flux vectors are non-local w.r.t. each other.
Nevertheless we will find that the ground state of the model is fully
${\cal N}=2$ supersymmetric.

For simplicity we consider the perturbative heterotic prepotential
with only three moduli $S$, $T$ and $U$ and $S\rightarrow\infty$:
$F=i(X^0)^2STU$. 
After the symplectic transformation
\begin{equation}
S\to F_S\,,\quad F_S\to -S\label{sfs}
\label{symptrans}
\end{equation}
the period vector has the form
\be
(\tilde X^I, \tilde F_I) = (1, -TU, iT, iU, -iSTU, iS, SU, ST) \;\;.
\label{PQsectiona}
\eq
One recognizes that the periods $\tilde X^I$ are algebraically dependent.
Now we choose the flux vectors such that all electric NS fluxes and also
all magnetic Ramond fluxes are zero, i.e.
 $e^1_I=m^{2I}=0$.
In addition,
in order to be able to balance the R fluxes against the NS fluxes we choose
them to be parallel, i.e. we impose the following condition:
\be
e_I^2/p=(l_2,-n_2,n_1,-l_1), \;\;\;m^{1I}/q=(-n_2,l_2,-l_1,n_1) \;\;.
\label{parallela}
\eq
In fact, these two flux vectors are mutually non-local,
$m \, \times \, e \, = \,
m^{1I}e^2_I-m^{2I}e^1_I \, = -2pq(l_1n_1+l_2n_2)\neq 0$.
With this choice the superpotential becomes
\be
W=(p+iqS\tau)(l_2-il_1U+in_1T-n_2UT),\label{supfac}
\eq
It is straightforward to see that the scalar potential
has a ${\cal N}=2$ supersymmetry preserving minimum with zero cosmological
constant;
specifically
the supersymmetry conditions,
\begin{equation}
W_S=W_T=W_U=W_\tau=0,\qquad {\rm and} \qquad W=0,
\end{equation}
have the following solution:
\begin{equation}
T_{\rm min}=\sqrt{{l_1l_2\over n_1n_2}},\quad U_{\rm min}=\sqrt{{l_2n_1\over
l_1n_2}},\quad S_{\rm min}~\tau_{\rm min}=i{p\over q}.
\end{equation}
So we see that supersymmetry preserving solutions are possible for finite
heterotic dilaton field $S$, as well 
as for finite type IIB dilaton $\tau$. Insisting on large $S$ and on large
$\tau$
implies that one has to ensure that $p\gg q$.

\section{Summary}

\setcounter{equation}{0}

In this paper we have provided a detailed investigation of the vacuum
structure of type IIA/B compactifications on Calabi-Yau spaces in
the presence of H-fluxes. These H-fluxes lift the vacuum degeneracy
in the Calabi-Yau moduli space. For aligned
fluxes, local minima of the
scalar potential with space-time supersymmetry and vanishing cosmological constant 
are found at several degeneration points of the Calabi-Yau space. 
However partial supersymmetry
at these minima seems to be impossible in local supergravity due
to the dilaton dependence of the effective superpotential, such that
the degeneration points always exhibit full ${\cal N}=2$ supersymmetry.
Away from the degeneration points supersymmetry is generically broken,
and we expect in general local minima of the scalar potential with
broken supersymmetry. We also examined the vacuum structure of some
Calabi-Yau spaces (quintic, sextic)
at certain rational, but non-singular points of enlarged
symmetry (Gepner points), and found that there are no supersymmetric
solutions at this points. However at the moment
we cannot completely exclude supersymmetric vacua at non-singular, rational
points for other Calabi-Yau spaces.

This discussion can be extended in several ways. For example it would
be interesting to map the type IIA/B H-fluxes to dual heterotic
or type I string compactifications. There the H-fluxes will correspond
to 
background electric or magnetic fields in the internal directions 
\cite{workprogress}\footnote{This was already conjectured for the flux
related to $X^0$ in \cite{PolStro}.} (for non-supersymmetric
string compactifications with background F-fields see
\cite{ffield}).
Similarly it would be nice to see \cite{workprogress} how
the type IIA/B H-fluxes at the Calabi-Yau singularities can be mapped
to dual brane configurations using the duality \cite{kls} between 
the Calabi-Yau geometrical engineering and the Hanany-Witten approach.

\section*{Acknowledgements}

We would like to thank Jan Louis for many valuable discussions and for
collaboration on the initial stage of this project. 
Discussions with K. Behrndt, G. Dell` Agata and G. Lopes-Cardoso
are also acknowledged.  
This work is supported by the European Commission RTN programme 
HPRN-CT-2000-00131 and by GIF - the German-Israeli Foundation
for Scientific Research. D.L. and S.T. thank the 
Erwin-Schroedinger-Institute in Vienna, where this project was started, for 
hospitality.

\section*{Appendix: The perturbative heterotic limit with general flux vector}

In this appendix we like to solve in
some detail the conditions of unbroken supersymmetry
within the perturbative heterotic limit for general flux vector
$(e_I,m^I)$.
The corresponding prepotential is given in eq.(\ref{hetprepot}).
After the symplectic transformation (\ref{symptrans})
the period vector has the form
\begin{equation}
(\tilde X^I;\tilde F_I)=(1,-T^a\eta_{ab}T^b,iT^a;-iST^a\eta_{ab}T^b,
iS,2S\eta_{ab}T^b).
\label{tildebasis}
\end{equation}
One recognizes that the periods $\tilde X^I$ are algebraically dependent.
The symplectic transformation  (\ref{sfs})
exchanges the IIA 2-cycle ${\cal C}_1^{(2)}\sim
vol(P^1_b)$ by its dual 4-cycle
${\cal C}_1^{(4)}$, namely the whole $K3$-fibre. 
This amounts in exchanging $H^{(4)}_R$ by $H^{(2)}_R$ 
in the $S$-field direction and
vice versa, and we denote the corresponding fluxes as
\begin{equation} 
\tilde e_1=m^1,\qquad \tilde m^1=e_1.
\end{equation}
Seen from the heterotic point of view, the symplectic basis
eq.(\ref{tildebasis}) is the most natural one since $\tilde X^I$ are
the perturbative periods associated to the electric $U(1)$,
whereas the $\tilde F_I$ are the non-perturbative magnetic $U(1)$
periods.

For simplicity  consider the case of only three moduli fields $S$, $T$ and
$U$.
The classical period vector is then
\be
(\tilde X^I, \tilde F_I) = (1, -TU, iT, iU, -iSTU, iS, SU, ST) \;\;.
\label{PQsection}
\eq
The holomorphic superpotential is then given as
\be
W= e_0 + \tilde e_1 TU + i e_2 T + i e_3 U + i m^0 STU + 
i \tilde m^1 S- m^2 SU - m^3 ST ,
\eq
and the K\"ahler potential take the well known form
\be
K = - \log\left[(S+\ov{S})(T+\ov{T})(U+\ov{U})\right] .
\eq
In order to keep the notation as simple as possible we will omit
from now on in this chapter the tilde on $e_1$ and $m^1$.

The perturbative duality transformations 
$SL(2, {\bf Z})_T \otimes
SL(2, {\bf Z})_U ) \times {\bf Z}_2^{T\leftrightarrow U}$
act as symplectic transformations on the period vector as
\be
\left( \begin{array}{c}
\tilde X^I \\ \tilde F_I \\
\end{array} \right) 
\rightarrow
\Gamma 
\left( \begin{array}{c}
\tilde X^I \\ \tilde F_I \\
\end{array} \right) = 
\left( \begin{array}{cc}
U & 0 \\ 0 & U^{T,-1} \\
\end{array} \right)
\left( \begin{array}{c}
\tilde X^I \\ \tilde F_I \\
\end{array} \right) \;\;.
\label{SympTrans}
\eq
The generalized K\"ahler function $e^G=e^K|W|^2$ 
is invariant under symplectic transformations
(\ref{SympTrans}) provided the quantum numbers are redefined
by $(m^I , -e_I) \rightarrow (m^I,-e_I) \Gamma^T$.
Note that under the modular transformations $SL(2, {\bf Z})_T \otimes
SL(2, {\bf Z})_U$ the superpotential transforms as a modular function
of modular weight -1, as required for the modular invariance of $G$
\cite{FLST}:
\begin{equation}
T\rightarrow{aT-ib\over icT+d}:\qquad W\rightarrow {W\over icT+d}.
\end{equation}

Now let us look for points of  which preserve
${\cal N}=1$ supersymmetry, i.e. we look for solutions of the
equation  
\beqa
D_a W = 0 \;\;  \rightarrow \;\; 
e^{K/2} \left(K_a W + W_a \right) =0 \;\;
, \;\; a = S, T, U  \;\;.\label{extrstu}
\eeqa
This problem was already solved 
\cite{CLM}  using  the attractor equations (\ref{susyeq}),
and the full solution of 
(\ref{extrstu}) 
is
\begin{eqnarray}
S_{\rm Susy} &=& i \frac{e \cdot m}{\la m,m \ra} 
+ \sqrt{ \frac{\la e ,e \ra }{\la m, m \ra}
- \frac{ (e \cdot m)^2 }{ \la m, m \ra^2 } } \;\;, \nonumber \\
T_{\rm Susy} &=& i \frac{e' \cdot m'}{\la m',m' \ra} 
+ \sqrt{ \frac{\la e' ,e' \ra }{\la m', m' \ra}
- \frac{ (e' \cdot m')^2 }{ \la m', m' \ra^2 } } \;\;,  \nonumber \\
U_{\rm Susy} &=& i \frac{e'' \cdot m''}{\la m'',m'' \ra} 
+ \sqrt{ \frac{\la e'' ,e'' \ra }{\la m'', m'' \ra}
- \frac{ (e'' \cdot m'')^2 }{ \la m'', m'' \ra^2 } } \;\;, 
\label{Solution}
\end{eqnarray}
where 
\begin{eqnarray} 
\la e, e \ra &=& 2 e_0 e_1 + 2 e_2 e_3\,, \nonumber\\
\la m, m \ra &=& 2 m^0 m^1 + 2 m^2 m^3\,,  \nonumber\\
e\cdot m &=&  e_0 m^0 + e_1 m^1 +  e_2 m^2 + e_3 m^3\,. 
\end{eqnarray}
The exchange symmetries $S\leftrightarrow T$ and 
$S \leftrightarrow U$ map the vectors $e,m$ to
vectors $e',m'$ and $e'', m''$.

The function $e^G$ at the supersymmetric point (\ref{Solution})
takes then the following value
\be
m_{3/2}^2|_{\rm Susy}=e^{G}|_{\rm Susy}= 
\sqrt{ \la e, e \ra \la m, m \ra - (e \cdot m)^2 }
= \la m, m \ra \, \Re S|_{\rm Susy} \;\;.
\label{ClassEntr}
\eq

{}From (\ref{ClassEntr}) 
one can easily read off that supersymmetric minima of the scalar potential
$v$ with $v=0$
at the minimum, i.e. $W|_{\rm Susy}=e^{G}|_{\rm Susy}=0$, are 
possible for certain
choices of fluxes.
The first possibility for solving these
two conditions is given by choosing parallel electric and magnetic
fluxes:
\be
e_I/p=(l_2,-n_2,n_1,-l_1), \;\;\;m^I/q=(-n_2,l_2,-l_1,n_1) \;\;.
\label{parallel}
\eq
This implies that
\beqa
\la e,  e \ra  &=& - 2 p^2 n^T l  \;\;, \nonumber\\
\la m, m \ra &=&  - 2 q^2 n^T l   \;\;, \nonumber\\
e \cdot m &=&   - 2 pq n^T l \;\;. \label{Hshort}
\eeqa
With this choice the superpotential is
\be
W=(p+iqS)(l_2-il_1U+in_1T-n_2UT),
\eq
and therefore $e^{G}|_{\rm Susy}=0$.
However for this class of solutions $\Re S$ is zero, so one is driven
to strong coupling which is not consistent with assumption of having
a large $S$-field.

The second class of supersymmetric
solutions is
given by 
only four non--vanishing flux
quantum numbers, namely the cases of purely electric ($m^I=0$) and
purely magnetic ($e_I=0$) charges. In both cases the moduli $T_{\rm Susy}$ and
$U_{\rm Susy}$ are generically finite.
For consistency  we have to require 
in addition
that
$S_{\rm min}=\infty$. This constraint is satisfied for the purely
electric solutions with $m^I=0$. 
This is precisely the case with aligned fluxes and
superpotential eq.(\ref{stualg}).
 On the other hand the  case with $e_I=0$
would imply strong coupling with $S_{\rm Susy}=0$.

Next let us consider the effects of one-loop corrections $h(T^a)$ to
the heterotic prepotential.

Due to the required embedding of the perturbative $T$-duality group
into the ${\cal N}=2$ symplectic transformations, it follows \cite{DKLL,AFGNT} 
that the heterotic one-loop prepotential $h(T^a)$ must 
obey well-defined transformation rules under this
group.
This becomes clear if one considers the action of the one-loop 
T-duality transformations on the period vector (\ref{tildebasis}):
\be
\left( \begin{array}{c}
\tilde X^I \\ i\tilde F_I \\
\end{array} \right) 
\rightarrow
\Gamma_{1-loop} 
\left( \begin{array}{c}
\tilde X^I \\ i\tilde F_I \\
\end{array} \right) = 
\left( \begin{array}{cc}
U & 0 \\ V & U^{T,-1} \\
\end{array} \right)
\left( \begin{array}{c}
\tilde X^I \\ i\tilde F_I \\
\end{array} \right) \;\;.
\label{SympTransa}
\eq
where the matrix $V$  encodes the quantum
corrections. 
Therefore the one-loop transformation rule of symplectic quantum numbers 
implied by the general formula $(m , -e) \rightarrow (m,-e) \Gamma^T$
is
\be
e \rightarrow U^{T,-1} e - V m,\;\;\;m \rightarrow U m \;\;.
\eq
It follows that the superpotential $W$ is still transforms
with modular weight -1 under $SL(2,{\bf Z})_T$ and $SL(2,{\bf Z})_U$.

In the presence of the heterotic one loop correction $h(T^a)$ the
period vector eq.(\ref{tildebasis}) will be 
modified as follows:
\begin{equation}
(\tilde X^I;\tilde F_I)|_{\rm 1-loop}
=(1,T^a\eta_{ab}T^b,iT^a;iS,iST^a\eta_{ab}T^b
+2ih(T^a)-iT^a{\partial h\over\partial T^a},-ST^a+{\partial h\over\partial T^a})
\label{tildebasismod}
\end{equation}
One recognizes, that the periods $\tilde X^I$ do not receive
any quantum corrections and are still algebraically dependent.
Hence  the superpotential with aligned fluxes is the same
as in the classical case eq.(\ref{stualg}), and  we
therefore expect the same vacuum structure as in the classical
limit for aligned fluxes. The one-loop quantum corrections only
result in a simple modification of the K\"ahler potential
which can be absorbed by the invariant dilaton field $S_{\rm invar}$ 
\cite{DKLL}.
This field $S_{\rm invar}$ is the true coupling constant at one loop,
and $e^{K_{\rm 1-loop}/2}$ is proportional to $S_{\rm invar}^{-1}$.

Let us consider in a little bit more
detail  the complete superpotential 
\begin{equation}
W=e_I\tilde X^I+m^I\tilde F_I|_{\rm 1-loop},
\end{equation}
which
includes the one-loop term $h(T^a)$ in $\tilde F_I|_{\rm 1-loop}$. 
The supersymmetry equations
$D_aW=0$, $W=0$ will still allow for non-trivial solutions which 
however cannot be expressed any more in closed form like in
eq.(\ref{Solution}) for non-vanishing $m^I$. 
Nevertheseless one can derive an all order expression for $m_{3/2}$
at the supersymmetric, stationary points, as it was proven in the
context of ${\cal N}=2$ black hole solutions.
This expression has simply the form
\be
m_{3/2}^2|_{\rm Susy}=
e^{G}|_{\rm Susy} =   \pi \Re S_{invar} \la m , m \ra \;\;,
\eq
where for the case of $N_V-1$ fields $T^a$,
$\la m,m\ra$ is defined as
\be
\la m,m\ra=m^0m^1+m^a\eta_{ab}m^b.
\eq
($\la m,m\ra$ is an invariant of the T-duality group $SO(2,N_V-1,{\bf Z})$.)
Thus the influence of all perturbative one-loop effects to the
superpotential is contained in $\Re S_{invar}$.
Hence, demanding $e^{G}|_{\rm min}=0$ at $\Re S_{invar}=\infty$,
one has to set again $\la m,m\ra =0$.

\end{document}